\documentclass[preprintnumbers,prd,showpacs,floatfix,superscriptaddress,nofootinbib,twocolumn]{revtex4-1}

\pdfoutput=1
\usepackage{graphicx,epsfig}
\usepackage{amsmath}
\usepackage {amssymb}
\usepackage {longtable}
\usepackage{multirow}

\usepackage{dcolumn}
\usepackage{bm}
\usepackage{amsfonts}
\usepackage{subfigure}
\usepackage{color}

\usepackage{relsize}

\def\be{\begin{equation}}
\def\ee{\end{equation}}
\def\beq{\begin{eqnarray}}
\def\eeq{\end{eqnarray}}

\begin{document}
\title{Junction conditions for generalized hybrid metric-Palatini gravity
with applications}





\author{Jo\~{a}o Lu\'{i}s Rosa}
\email{joaoluis92@gmail.com}
\affiliation{Institute of Physics, University of Tartu, W. Ostwaldi
1, 50411 Tartu, Estonia}

\author{Jos\'{e} P. S. Lemos}
\email{joselemos@ist.utl.pt}
\affiliation{Departamento de F\'isica,
Centro de Astrofisica e Gravita\c c\~ao
- CENTRA,
Instituto Superior T\'{e}cnico - IST,
Universidade de Lisboa - UL,
Avenida Rovisco Pais 1, 1049-001, Portugal}


\begin{abstract}

The generalized hybrid metric-Palatini gravity is a theory of
gravitation that has an action composed of a Lagrangian given by
$f(R,\cal R)$, where $f$ is a function of the metric Ricci scalar $R$
and a new Ricci scalar $\cal R$ formed from a Palatini connection,
plus a matter Lagrangian.  This theory can be rewritten by trading the
new geometric degrees of freedom that appear in $f(R,\cal R)$ into two
scalar fields, $\varphi$ and $\psi$, yielding a dynamically equivalent
scalar-tensor theory.  Given a spacetime theory, the next important
step is to find solutions within it. To construct appropriate
solutions it is often necessary to know the junction conditions
between two spacetime regions at a separation hypersurface
$\Sigma$, with each
spacetime region being an independent solution of the theory. The
junction conditions for the generalized hybrid metric-Palatini gravity
are found here, both in the geometric representation and in the
scalar-tensor representation, and in addition, for each
representation, the junction conditions for a matching with a
thin-shell of matter and for a smooth matching at the separation
hypersurface are worked out.  These junction conditions are then
applied to three configurations, namely, a star, a quasistar with a
black hole, and a wormhole.  The star is made of a Minkowski interior,
a thin shell at the interface with all the matter energy conditions
being satisfied, and a Schwarzschild exterior with mass $M$, and
unlike general relativity where the matching can be performed at any
radius $r_\Sigma$, for this theory the matching can only be performed
at a specific value of the shell radius, namely $r_\Sigma=\frac{9M}4$,
that corresponds to the general relativistic Buchdahl radius.  The
quasistar with a black hole is made of an interior Schwarzschild black
hole surrounded by a thick shell that matches smoothly to a mass $M$
Schwarzschild exterior at the light ring radius $r_{\Sigma e}=3M$, and
with the matter energy conditions being satisfied for the whole
spacetime. The wormhole is made of some interior with matter that
contains the throat, a thin shell at the interface, and a
Schwarzschild-AdS exterior with mass $M$ and negative cosmological
constant $\Lambda$, with the matter null energy condition being obeyed
everywhere within the wormhole.

\end{abstract}

\maketitle

\section{Introduction}\label{introduction}
\subsection{Junction conditions in theories of gravitation and
applications}

In field theories in general, and in particular in theories of
gravitation, one has to find the theory's junctions conditions that
match, through a matching surface, a solution in a given region to
another solution in a neighbor region.  An example is in Newtonian
gravitation where, for a boundary surface it is imposed that the
gravitational potential and its first derivatives are continuous
across the surface, and for a surface layer, i.e.,
an infinitesimally
thin shell, it is imposed that the
gravitational potential and its first derivatives on the surface are
continuous but its first derivative normal to the surface is
discontinuous yielding in a simple case the shell's mass density, the
shell's pressure being then found through the equation of motion.

General relativity has its own junction conditions.  Indeed, to find
solutions of the Einstein's field equations it is often necessary to
match two solutions each defined in a given region that join at some
hypersurface. The whole spacetime is thus described by two or more
regions with different metrics tensors expressed in different
coordinate systems.  The junction conditions in general relativity for
a boundary surface were found by Darmois \cite{darmois} by imposing
that the induced metric across the hypersurface that separates the two
spacetime regions must be continuous, and in addition the extrinsic
curvature of that hypersurface must also be continuous.  Lichnerowicz
\cite{lichnerowicz} also gave a set of conditions to have at a
junction which are coordinate dependent, see \cite{lake} for a
comparison between both sets of conditions.  For a surface layer,
i.e., a thin-shell, Lanczos had a first go at this junction
condition problem \cite{lanczos1},
which was then picked up by Sen that
in matching an interior Minkowski
to an exterior Schwarzschild spacetime
found, in the context of a thin shell
the critical radius that appears in the interior
Schwarzschild solution, now called the Buchdahl radius
\cite{sen}.
There were further developments by 
 Lanczos himself  \cite{lanczos2}
 and then the formalism 
was definitely closed when  Israel
put into a Gauss-Codazzi system of
equations \cite{Israel:1966rt}, showing that if the
extrinsic curvature is not continuous, then the matching between two
regions of spacetime can still be done, but implies the existence of a
thin-shell of matter at the junction radius.  The thin shell
problem was also
studied in \cite{papa}, and  Taub gave a very general formalism
of which the thin shell fits into \cite{taub}.
Of course, each theory
of gravitation has its own junction conditions which must be deduced
from the complete set of field equations.  The first modified theory
of gravitation
put forward was general relativity with a cosmological constant term
and this theory
has the same junction conditions as general relativity itself.
In several other theories the junction conditions have been
derived. For the Einstein-Cartan theory see \cite{firstattemptEC}, for
the Brans-Dicke theory and other scalar-tensor theories see
\cite{suffern,Barrabes:1997kk}, for Gauss-Bonnet gravity see
\cite{Davis:2002gn}, for quadratic gravity theories, such as quadratic
in the Riemann tensor, Ricci tensor or Ricci scalar $R$, see
\cite{Senovilla:2014kua,berezin}, for
theories with a gravitational Lagrangian density
that is a function $f$ of $R$, i.e.,
$f\left(R\right)$ theories,
without torsion see \cite{Deruelle:2007pt,Senovilla:2013vra} and with
torsion see \cite{Vignolo:2018eco},
for the hybrid metric-Palatini gravity,
which is a 
development of $f\left(R\right)$ theories
now with Lagrangian density
$R+f(\cal R)$, where $\cal R$ is a
new Ricci scalar derived from a new
connection, see \cite{olmo}, and for  
another
development of $f\left(R\right)$ theories,
namely, theories where the Lagrangian
density depends on a function of
both $R$ and the trace of the stress-energy tensor
$T$, i.e., $f\left(R,T\right)$ theories, see \cite{Rosa:2021teg}.

The junction conditions have a great many number of applications as
they are used to derive new solutions and so yield new insights of the
corresponding theory of gravitation.  In general relativity there are
applications to star solutions without and with thin shells and as
models for black hole mimickers see \cite{lemoszachin,rosafluid}, to
gravitational collapse with different interiors and exteriors
\cite{senovilla1} and gravitational collapse of the Oppenheimer-Snyder
in many different coordinate systems, see e.g., \cite{silva},
to quasistars,
i.e., matter surrounded a black hole that shines due to the black hole
gravitational field \cite{frauendiener}, and to wormholes where
junctions between one side of the universe and the other side can be
performed \cite{lemoswormhole1,lemoswormholes2}.  In Einstein-Cartan
theory there are applications of junction conditions to spacetimes
containing compact objects \cite{luzcarloni}, in Brans-Dicke theory
there are applications to wormhole solutions \cite{sushkov}, in
Gauss-Bonnet gravity there are applications of thin shells in brane
worlds \cite{ramirez}, in quadratic gravity there are applications to
wormholes \cite{eiroa}, in $f(R)$ theories of gravitation there are
applications involving thin shell stars \cite{eiroa2}, in
the hybrid metric-Palatini extension to $f(R)$ gravity theories
there are applications to wormholes
\cite{bronnikovhybrid},
and in
the other extension to $f(R)$
that
includes the trace of the stress-energy tensor
there are applications to stars
\cite{maurya},
to name a few.

\subsection{Junction conditions in the generalized hybrid
metric-Palatini gravity and applications}

The
generalized hybrid metric-Palatini gravity is a theory of gravitation
that generalizes
the hybrid
metric-Palatini Lagrangian density $R+f(\cal R)$
into a generic function $f(R,\cal R)$ \cite{Tamanini:2013ltp},
so that the generalized hybrid metric-Palatini gravity
theory is an extra development of 
$f(R)$ gravity theories.
The $f(R,\cal R)$ theory can be written in two representations.
One is the geometric representation given by the
function $f(R,\cal R)$ itself. 
The other is the scalar-tensor representation where 
some of the degrees of freedom
in the function $f$ of the geometric fields $R$ and $\cal R$
can be traded 
for  two scalar fields, yielding a dynamically equivalent
scalar-tensor theory, in the same manner as the $f\left(R\right)$
theory can be rewritten as a dynamically equivalent scalar-tensor
theory with one scalar field.  The generalized hybrid metric-Palatini
gravity has very interesting features and has been studied in the
context of cosmological solutions
\cite{Rosa:2017jld},
wormholes \cite{rosa2018},
scalar modes.
\cite{bombacigno},
dynamical systems
\cite{rosa2019},
stability of Kerr
black holes \cite{rosabh},
cosmic strings \cite{Rosa:2021zbk},
thick-brane structures \cite{rosatb},
singularities in cosmological models 
\cite{Rosa:2021ish},
weak-field regime \cite{Rosa:2021lhc},
and 
double layers applied to wormholes
\cite{Rosa:2021yym}.
Junction conditions must be found for any theory of gravitation,
including the generalized hybrid metric-Palatini gravity.

In this work, we deduce the junction conditions of the generalized
hybrid metric-Palatini gravity.  We display these conditions in both
the geometrical and the scalar-tensor representations.
For each representation the junction
conditions for a matching with a thin-shell of matter and for a smooth
matching at the separation hypersurface are worked out.
We show three
applications: stars, quasistars with black holes, and wormholes.

The paper is organized as follows.
In Sec.~\ref{equations1}, we display de action of the theory and the
field equations both in the geometrical and in the scalar-tensor
representations.
In Sec.~\ref{geometrical}, we obtain the sets of junction conditions in
the geometrical and in the scalar-tensor representations and in each
representation we give the conditions for thin shell matching and for
smooth matching.
In Sec.~\ref{calculationsforsphericalthinshells}, we
present the junction conditions for static spherically symmetric
spacetimes.
In Sec.~\ref{app}, the first application of the junction conditions is
made to a star thin shell, more precisely, for an inner Minkowski
spacetime, a thin shell in the middle, and an outer Schwarzschild
spacetime.
In Sec.~\ref{secondapplication}, the second application is made to a
quasistar with a black hole, it is an application involving smooth
matching, an inner Schwarzschild black hole, a mild thin shell at the
inner junction, followed by a thick shell that matches smoothly into
an outer Schwarzschild vacuum spacetime.
In Sec.~\ref{thirdapplication}, the third application is presented, to
a wormhole involving a thin shell and an outer AdS
spacetime.
In Sec.~\ref{conclusion}, we conclude.
The Appendices~\ref{choosingatheory} and~\ref{calculationsappendixb}
are used as auxiliary
tools to the main text.

\section{Equations of the
generalized hybrid metric-Palatini gravity theory}
\label{equations1}

\subsection{Equations of
the geometrical representation of the theory}

Consider the generalized hybrid metric-Palatini gravity
action $S$ given by
\be\label{genac}
S=\frac{1}{16\pi}\int_\Omega\sqrt{-g}f\left(R,\cal{R}\right)d^4x+S_m,
\ee
where we have chosen a system of geometrized units
for which the gravitational constant $G$
and the speed of light $c$ are set to one,
$\Omega$ is the entire region
of the spacetime manifold, 
$g$ is the determinant of the spacetime
metric $g_{ab}$, 
$R$ is the metric Ricci scalar,
$\mathcal{R}$ is the Palatini Ricci
scalar,
defined by 
$\mathcal{R}\equiv\mathcal{R}^{ab}g_{ab}$,
where the the Palatini Ricci tensor $\mathcal{R}^{ab}$
is defined in terms of an
independent connection $\hat\Gamma^c_{ab}$ as, 
$
\mathcal{R}_{ab}=\partial_c
\hat\Gamma^c_{ab}-\partial_b\hat\Gamma^c_{ac}+
\hat\Gamma^c_{cd}\hat\Gamma^d_{ab}-\hat\Gamma^c_{ad}\hat\Gamma^d_{cb}
$, $f\left(R,\cal{R}\right)$ is a well behaved function of $R$ and
$\cal{R}$, $S_m$ is the matter action defined as $S_m=\int
d^4x\sqrt{-g}\;{\cal L}_m$ and where ${\cal L}_m$ is the matter
Lagrangian density considered minimally coupled to the metric
$g_{ab}$.
Variation of the action (\ref{genac}) with respect to the
metric $g_{ab}$ yields the following equation of motion
$
f_R
R_{ab}+f_{\cal R}\mathcal{R}_{ab}-\frac{1}{2}
g_{ab}f\left(R,\cal{R}\right)
-\left(\nabla_a\nabla_b-g_{ab}\Box\right)f_R
=8\pi T_{ab}
$,
where, $f_R\equiv\frac{\partial f}{\partial R}$, 
$f_{\cal R}\equiv\frac{\partial f}{\partial {\cal R}}$,
with $f\equiv f\left(R,\cal{R}\right)$,
$\nabla_a$ is the  $\Gamma^c_{ab}$
metric connection
covariant derivative,
with $
\Gamma^a_{bc}=\frac{1}{2}{g}^{ad}
\left(\partial_b {g}_{dc}+\partial_c
{g}_{bd}-\partial_d {g}_{bc}\right)
$,
$\Box=\nabla^a\nabla_a$ is the d'Alembertian,
 and
$T_{ab}$ is the stress-energy tensor defined as
$
T_{ab}=-\frac{2}{\sqrt{-g}}\frac{\delta(\sqrt{-g}\,{\cal
L}_m)}{\delta(g^{ab})}
$.
Varying the action (\ref{genac}) with respect to the independent
connection $\hat\Gamma^c_{ab}$ provides the following
equation
$
\hat\nabla_c\left(\sqrt{-g}
f_{\cal{R}}
g^{ab}\right)=0 
$.
where $\hat\nabla_a$ is the $\hat\Gamma^c_{ab}$
connection
covariant derivative.
Now, for
the scalar density $\sqrt{-g}$
of weight 1, we
have that $\hat\nabla_c \sqrt{-g}=0$ and so the latter
equation 
simplifies to $\hat\nabla_c\left(
f_{\cal{R}}\,
g^{ab}\right)=0$.
This means that ${\hat g}_{ab}$, defined as
$
{\hat g}_{ab}=f_{\cal{R}}\,g_{ab} 
$,
is a metric compatible with the connection $\hat\Gamma^a_{bc}$
which then can be written as the following Levi-Civita connection
$
\hat\Gamma^a_{bc}=\frac{1}{2}{\hat g}^{ad}
\left(\partial_b {\hat g}_{dc}+\partial_c
{\hat g}_{bd}-\partial_d {\hat g}_{bc}\right)
$
where $\partial_a$ denotes partial derivative.  Note also
that ${\hat g}_{ab}$ is conformally related to $g_{ab}$
through the conformal factor $f_{\cal{R}}$. This
result implies that the two Ricci tensors $R_{ab}$ and $\mathcal
R_{ab}$, that we assumed to be independent at first, are actually
related to each other by
$
\mathcal R_{ab}=R_{ab}-\frac{1}{f_\mathcal
R}\left(\nabla_a\nabla_b+\frac{1}{2}g_{ab}\Box\right)f_\mathcal
R+\frac{3}{2f_\mathcal R^2}\partial_a f_\mathcal R\partial_b
f_\mathcal R
$,
where again $\Box=\nabla^c\nabla_c$.
This relation allows us to eliminate
$\mathcal R_{ab}$
from the equation obtained from
variation of the action (\ref{genac}) with respect to the
metric $g_{ab}$, see above, to get
\beq\label{genfield2}
&&\left(R_{ab}-\nabla_a\nabla_b+g_{ab}\Box\right)\left(f_R+
f_\mathcal R\right)-\nonumber \\
&&-\frac{3}{2}g_{ab}\Box f_\mathcal R+\frac{3}{2f_\mathcal R}
\partial_af_\mathcal R\partial_bf_\mathcal R-
\frac{1}{2}g_{ab}f=8\pi T_{ab}\,.
\eeq
The other  equation of motion obtained from 
varying the action (\ref{genac}) with respect to the independent
connection $\hat\Gamma^c_{ab}$, namely, 
$
\hat\nabla_c\left(\sqrt{-g}
f_{\cal{R}}
g^{ab}\right)=0 
$, see above, can then be swapped 
by the equation that relates the
two Ricci tensors $R_{ab}$ and $\mathcal
R_{ab}$, namely, 
\be\
\label{riccirel}
\mathcal R_{ab}=R_{ab}-\frac{1}{f_\mathcal
R}\left(\nabla_a\nabla_b+\frac{1}{2}g_{ab}\Box\right)f_\mathcal
R+\frac{3}{2f_\mathcal R^2}\partial_a f_\mathcal R\partial_b
f_\mathcal R\,,
\ee
see above.
Considering that $f$ is a function of the two variables $R$ and
$\mathcal R$, we can write the partial derivatives $\partial_a f_X$,
and the covariant derivatives $\nabla_a\nabla_bf_X$, with $X$ being
either
$R$ or $\mathcal R$, as
\beq\label{partialf}
&&\partial_a f_X= f_{XR}\partial_aR+f_{X\mathcal R}
\partial_a\mathcal R\,,
\nonumber\\
&&\nabla_a\nabla_bf_X=f_{XR}\nabla_a\nabla_bR+f_{X\mathcal R}
\nabla_a\nabla_b\mathcal R+\label{ppartialf}\\
&&+f_{XRR}\partial_aR\partial_bR+f_{X\mathcal R\mathcal R}
\partial_a\mathcal R\partial_b\mathcal R+
2f_{XR\mathcal R}\partial_{(a}R\partial_{b)}\mathcal R\,.
\nonumber
\eeq
These results allow us to expand
the terms with derivatives of $f_R$ or $f_\mathcal R$ in
Eq.~\eqref{genfield2} and write them as derivatives of either $R$ or
$\mathcal R$. We do not write the resultant equation here due
to its size,
it can be found in
Appendix~\ref{longformula}.

\subsection{Equations of
the scalar representation of the theory}

It is sometimes useful to express the action \eqref{genac} in a
scalar-tensor representation. This can be achieved by considering an
action with two auxiliary fields, $\alpha$ and $\beta$, respectively,
in the following form
$
S=\frac{1}{16\pi}\int_\Omega \sqrt{-g}\Big[f\left(\alpha,
\beta\right)+\frac{\partial f}{\partial \alpha}\left(R-\alpha\right)
+\frac{\partial f}{\partial\beta}\left(\cal{R}-\beta
\right)\Big]d^4x+S_m
$.
Using $\alpha=R$ and $\beta=\mathcal{R}$ we recover action
\eqref{genac}. Therefore, we can define two scalar fields as
$\varphi=\frac{\partial
f}{\partial\alpha}$
and
$\psi=-\frac{\partial
f}{\partial\beta}$, where the negative sign is put here for
convention. The equivalent action is of the form
$
S=\frac{1}{16\pi}\int_\Omega \sqrt{-g}
\left[\varphi R-\psi\mathcal{R}-V\left(\varphi,\psi\right)\right]d^4x
$,
where we defined the potential $V\left(\varphi,\psi\right)$ as
$
V\left(\varphi,\psi\right)=-f\left(\alpha,\beta\right)+
\varphi\alpha-\psi\beta
$.
Taking into account that ${\hat g}_{ab}$
is conformally related to $g_{ab}$ through the relation 
$
{\hat g}_{ab}=f_{\cal{R}}\,g_{ab} 
$,
and that it
can now be written as
${\hat g}_{ab}=-\psi\, g_{ab}$, we have that the Ricci scalars
$R$ and $\mathcal{R}$ are related through
$
\mathcal{R}=R+\frac{3}{\psi^2}\partial^a \psi\partial_a \psi-
\frac{3}{\psi}\Box\psi
$.
So, we can replace $\cal{R}$ into the action just derived, 
to obtain
\beq\label{genacts2}
S=\frac{1}{16\pi}\int_\Omega \sqrt{-g}\Big[
\left(\varphi-\psi\right) R-
\frac{3}{2\psi}\partial^a\psi\partial_a\psi
     \nonumber  \\
-V\left(\varphi,\psi\right)\big]d^4x+S_m\,,
\eeq
where $S_m$ is the matter action defined before.
Varying the action \eqref{genacts2} with respect
to the metric $g_{ab}$
provides the following gravitational equation 
\beq
&&\left(\varphi-\psi\right) G_{ab}
-\nabla_a\nabla_b\varphi
-\frac{3}{2\psi}\partial_a\psi\partial_b\psi
+\nabla_a\nabla_b\psi+
	\nonumber  \\
&&+\left(\Box\varphi-\Box\psi
+\frac{3}{4\psi}
\partial^c\psi\partial_c\psi
+\frac{1}{2}V
\right)g_{ab}=8\pi T_{ab}\,.
\label{genein2}
\eeq
Varying the action with respect to the field $\varphi$ and to the 
field $\psi$ yields, after rearrangements,
\be\label{genkgi}
\Box\varphi+\frac{1}{3}\left(2V-\psi V_\psi-\varphi
V_\varphi\right)=\frac{8\pi T}{3}\,,
\ee
and
\be\label{genkg}
\Box\psi-\frac{1}{2\psi}\partial^a\psi\partial_a\psi-
\frac{\psi}{3}\left(V_\varphi+V_\psi\right)=0\,,
\ee
respectively.

\section{Junction conditions for the generalized hybrid
metric-Palatini gravity theory}
\label{geometrical}

\subsection{Junction conditions for 
the geometrical representation of the theory}

\subsubsection{Matching with a thin-shell
at $\Sigma$}\label{sec:geomat}

Let us denote the whole four-dimensional spacetime by $\mathcal V$,
which is divided by a hypersurface $\Sigma$ into two regions,
$\mathcal V^+$ and $\mathcal V^-$.  We denote the metric $g_{ab}^+$,
given in coordinates $x^a_+$, as the metric in region $\mathcal V^+$,
and the metric $g_{ab}^-$, given in coordinates 
$x^a_-$, as the metric
in region $\mathcal V^-$, with latin indices running from $0$ to $3$,
$0$ being in general a time index.  In both sides of $\Sigma$, we
define a set of coordinates $y^\alpha$, with greek indices running
from $0$ to $2$, or some other combination of three indices out of the
four latin indices.  We define the projection vectors from the
four-dimensional regions  $\mathcal V^+$ and $\mathcal V^-$, to the
three-dimensional hypersurface $\Sigma$ as $e^a_\alpha=\frac{\partial
x^a}{\partial y^\alpha}$.  The unit normal vector $n^a$ to $\Sigma$ is
defined to point in the direction from $\mathcal V^-$ to $\mathcal
V^+$.  We denote by $l$ the proper distance or proper time along the
geodesics perpendicular to $\Sigma$. The parameter $l$ is chosen equal
to zero at $\Sigma$, is positive in the region $\mathcal V^+$, and is
negative in the region $\mathcal V^-$.  The infinitesimal displacement
along the geodesics is $dx^a=n^adl$, with $l$ parameterizing the
geodesic and the normal $n_a$ is here given by $n_a=\epsilon
\partial_a l$, with $\epsilon$ being either $-1$ or $+1$ for $n^a$ a
timelike or spacelike vector, respectively, so $n_a$ satisfies
$n^an_a=\epsilon$.  To match the two regions $\mathcal V^+$ and
$\mathcal V^-$ through a hypersurface $\Sigma$, the distribution function
formalism is in general needed, so we define $\Theta\left(l\right)$ as
the Heaviside distribution function, and
$\delta\left(l\right)=\partial_l\Theta\left(l\right)$ as the Dirac
distribution function.  For a quantity $X$, we write
$X=X^+\Theta\left(l\right)+X^-\Theta\left(-l\right)$, where the index
$+$ indicates that the quantity $X^+$ is the value of the quantity $X$
in the region $\mathcal V^+$, and the index $-$ indicates that the
quantity $X^-$ is the value of the quantity $X$ in the region
$\mathcal V^-$.  The jump of $X$ across $\Sigma$ is denoted by
$\left[X\right]=X^+|_\Sigma-X^-|_\Sigma$.  The normal $n^a$ and the
tangent vector $e^a_\alpha$ to $\Sigma$ have zero jump by definition,
i.e., $\left[n^a\right]=0$ and $\left[e^a_\alpha\right]=0$.

We now derive the junction conditions for the geometrical
representation of the generalized hybrid metric-Palatini gravity. We
deal with $g_{ab}$ to start with and only after with
$\hat\Gamma^a_{bc}$.  We consider the general case for which a
thin-shell arises at the matching hypersurface.

Let us start with $g_{ab}$.
To have a spacetime with a line element, and so a metric $g_{ab}$,
this has to be properly defined throughout
the spacetime. In particular, the metric
must have some form of continuity.
In the distribution formalism, one writes the metric $g_{ab}$ as
\be
g_{ab}=g_{ab}^+\Theta(l)+g_{ab}^-\Theta(-l)\,.
\label{metricsplit}
\ee
The derivative of $g_{ab}$ is
$\partial_c g_{ab}=
(\partial_c g_{ab}^+)\Theta(l)+(\partial_c g_{ab}^-)\Theta(-l)
+\epsilon 
\left[g_{ab}\right]n_c\delta(l)
$.
The term proportional to $\delta\left(l\right)$ is problematic, 
because the Christoffel symbols corresponding to it
would have products of the form
$\Theta\left(l\right)\delta\left(l\right)$ which are not defined in
the distribution formalism. Therefore one has
to impose $\left[g_{ab}\right]=0$.
Moreover, generically
$g_{ab}$ induces a metric on 
$\Sigma$ which is given $h_{\alpha\beta}=g_{ab}e^a_\alpha
e^b_\beta$, such that from the exterior the induced metric is
$h_{\alpha\beta}^+=g_{ab}^+e^a_\alpha
e^b_\beta$
and from the interior the induced metric is
$h_{\alpha\beta}^-=g_{ab}^-e^a_\alpha
e^b_\beta$.
So, for $h_{\alpha\beta}$ to give a continuous metric
on $\Sigma$ we must have $h_{\alpha\beta}^+-h_{\alpha\beta}^-=0$, i.e.,
\be
\label{junction1a}
\left[h_{\alpha\beta}\right]=0.
\ee
This junction
condition is the same as the
first junction condition in general relativity, and also should hold
generically for many theories of gravitation.
Then the derivative of the metric is now
\be
\partial_c g_{ab}=
(\partial_c g_{ab}^+)\Theta(l)+(\partial_c g_{ab}^-)\Theta(-l)
\,.
\ee
Now, let us analyze 
the further junction
conditions related to $g_{ab}$ that arise in the theory.
Notice that Eq.~\eqref{genfield2} depends directly on the
function $f$, which can be any general function of $R$ and $\mathcal
R$. This means that, in general, there will be terms in $f$ that are
power-laws or products of $R$ and $\mathcal R$. When we write these
terms as distribution functions, extra junction conditions will arise
in order to prevent the appearance of terms of the form
$\Theta\left(l\right)\delta\left(l\right)$, undefined in the
distribution formalism.
Let us analyze first the zero order derivative term, $R$.
In general, the Ricci tensor $R_{ab}$
of the metric $g_{ab}$
can be written in terms of distribution functions as
$
R_{ab}=R_{ab}^+\Theta\left(l\right)+
R_{ab}^-\Theta\left(-l\right)
-\big(\epsilon\,
e_a^\alpha e_b^\beta
\left[K_{\alpha\beta}\right]+n_an_b\left[K\right]\big)
\delta\left(l\right)$, 
where
$K_{\alpha\beta}=\nabla_\alpha n_\beta$
is the extrinsic curvature of $\Sigma$
with $n_\beta=e_\beta^b n_b$, and 
$K=K^\alpha_\alpha$ is
the trace of $K_{\alpha\beta}$.
The Ricci
scalar is then 
$R=R^+\Theta\left(l\right)+R^-\Theta\left(-l\right)-
2\epsilon\left[K\right]
\delta\left(l\right)$.
To avoid the presence of singular terms of the form
$\delta\left(l\right)^2$ in the products between Ricci scalars, say,
in the
function $f$, a junction condition arising from this analysis is
\be\label{junction1b}
\left[K\right]=0.
\ee
This condition does not appear in general relativity.
Using Eq.~(\ref{junction1b}) the Ricci
 tensor is 
\be\label{riccitensor}
R_{ab}=R_{ab}^+\Theta\left(l\right)+
R_{ab}^-\Theta\left(-l\right)
-\epsilon\left[K_{ab}\right]
\delta\left(l\right)\,,
\ee
and the Ricci
scalar is now
\be\label{distric}
R=R^+\Theta\left(l\right)+R^-\Theta\left(-l\right)\,.
\ee
Let us analyze now the first
order derivative term $\partial_a R$.
Computing the partial derivatives of $R$  expressed in
Eq.~\eqref{distric} leads to
$\partial_a R=\partial_aR^+\Theta\left(l\right)+
\partial_aR^-\Theta\left(-l\right)+\epsilon
\left[R\right]
n_a\delta\left(l\right)$.
In the field equations in Eq.~\eqref{genfield2}, we can see that due
to the existence of the term $\partial_af_\mathcal
R\partial_bf_\mathcal R$, there will be terms depending on products of
these derivatives, such as $\partial^cR\partial_cR$.
Given the terms that appear in $\partial_a R$ these
products would depend on $\delta\left(l\right)^2$, which are singular
terms, or on $\Theta\left(l\right)\delta\left(l\right)$, which are
undefined. Therefore, to avoid the presence of these terms we obtain
a junction condition for $R$ as
\be\label{juncric}
\left[R\right]=0,
\ee
Then $\partial_a R$ can be written as
\be
\partial_a R=\partial_aR^+\Theta\left(l\right)+
\partial_aR^-\Theta\left(-l\right)\,.
\label{derivric}
\ee
Form Eq.~(\ref{juncric}) we see that
$R$ must be continuous across the hypersurface
$\Sigma$. We then denote the value of $R$ 
at $\Sigma$ as $R_\Sigma$.

Let us now turn to the independent connection $\hat\Gamma$. As
we have seen, the Palatini Ricci tensor $\mathcal
R_{ab}$ is written in terms of $\hat\Gamma$ and its
derivatives,
$\mathcal{R}_{ab}=\partial_c
\hat\Gamma^c_{ab}-\partial_b\hat\Gamma^c_{ac}+
\hat\Gamma^c_{cd}\hat\Gamma^d_{ab}-\hat\Gamma^c_{ad}\hat\Gamma^d_{cb}$.
Consequently, the Palatini Ricci scalar $\mathcal R$ will
also depend on $\hat\Gamma$ in the same way. As $\mathcal R$ is
generally present in the field equations through the
function $f\left(R,\mathcal R\right)$ and its derivatives,
extra junction conditions
will arise from $\hat\Gamma$. Being a fundamental field of the theory,
$\hat\Gamma$ can be written in the distribution formalism as
\be
\hat\Gamma^c_{ab}=\hat\Gamma^{c+}_{ab}
\Theta\left(l\right)+\hat\Gamma^{c-}_{ab}\Theta\left(-l\right)\,,
\ee
where $\delta$ terms are not present to avoid the presence
of undefined terms.
Defining the extrinsic curvature written in terms of the independent
connection $\hat\Gamma$ as $\mathcal K_{ab} = e_a^\alpha
e_b^\beta\hat\nabla_\alpha n_\beta$,
where $\hat\nabla$ is the covariant derivative on the
boundary with respect to $\hat\Gamma$, 
the trace  of $\mathcal K_{ab}$ as $\mathcal K$, and
following the previous arguments for $R_{ab}$ and $R$ that led
Eq.~\eqref{junction1b}, we obtain
\be\label{junction1b2}
\left[\mathcal K\right]=0\,.
\ee
Then, the Ricci tensor is
\be\label{riccitensorcalig}
\mathcal R_{ab}=\mathcal R_{ab}^+\Theta\left(l\right)+
\mathcal R_{ab}^-\Theta\left(-l\right)
-\epsilon\left[{\mathcal K}_{ab}\right]
\delta\left(l\right)\,.
\ee
and with the help of Eq.~(\ref{junction1b2}) the corresponding
Ricci scalar is
\be\label{distmric}
\mathcal R=\mathcal R^+\Theta\left(l\right)+
\mathcal R^-\Theta\left(-l\right).
\ee
Let us now work with $\partial_a \mathcal R$.
Computing the partial derivatives of $\mathcal R$ expressed in
Eq.~\eqref{distmric} leads to
$\partial_a \mathcal R=\partial_a\mathcal R^+\Theta\left(l\right)+
\partial_a\mathcal R^-\Theta\left(-l\right)+
\epsilon\delta\left(l\right)\left[\mathcal R\right]n_a$.
In the field equations in Eq.~\eqref{genfield2}, we can see that due
to the existence of the term $\partial_af_\mathcal
R\partial_bf_\mathcal R$, there will be terms depending on products of
these derivatives, such as $\partial^c{\mathcal R}\partial_c{\mathcal
R}$.  Given the terms that appear in $\partial_a \mathcal R$,
again these products
would depend on $\delta\left(l\right)^2$, which are singular terms, or
on $\Theta\left(l\right)\delta\left(l\right)$, which are
undefined. Therefore, to avoid the presence of these terms we obtain
the junction conditions for $\mathcal R$ as
\be\label{juncriccal}
\left[\mathcal R\right]=0.
\ee
Then $\partial_a \mathcal R$ can be written as
\be
\partial_a \mathcal R=\partial_a\mathcal R^+\Theta\left(l\right)+
\partial_a\mathcal R^-\Theta\left(-l\right)\,.
\label{derivric2}
\ee
Form Eq.~(\ref{juncriccal}) we see that
$\mathcal R$ must be continuous across the hypersurface
$\Sigma$. We then denote the value of $\mathcal R$ 
at $\Sigma$ as $\mathcal R_\Sigma$.
Clearly, 
Eqs.~\eqref{juncric} and~\eqref{juncriccal},
imply that the terms with first derivatives of $R$ and
$\mathcal R$, see Eqs.~(\ref{derivric}) and~(\ref{derivric2})
are regular.

We now turn to the jumps of the derivatives of $R$ and $\mathcal R$.
These jumps are not independent of each other, and the relationship
between them can be obtained from Eq.~\eqref{riccirel}. To find this
relationship, let us first write the second order derivatives of $R$
and $\mathcal R$, i.e., the terms $\nabla_a\nabla_bR$, $\Box R$,
$\nabla_a\nabla_b \mathcal R$, and $\Box\mathcal R$, in the
distribution formalism. The second order term $\nabla_a\nabla_bR$ can
be generically written in the distribution formalism as
\begin{eqnarray}
\nabla_a\nabla_bR&=\nabla_a\nabla_bR_+\Theta\left(l\right)+
\nabla_a\nabla_bR_-\Theta\left(-l\right)+\nonumber \\
&
\epsilon n_a\left[\partial_bR\right]\delta\left(l\right)\,,
\label{second1}
\end{eqnarray}
and then
$\Box R=\Box R_+\Theta\left(l\right)+
\Box R_-\Theta\left(-l\right)+
n^a\left[\partial_aR\right]\delta\left(l\right)$.
Likewise, 
the
second order term $\nabla_a\nabla_b{\mathcal R}$
can be generically written in the distribution formalism as
\begin{eqnarray}
\nabla_a\nabla_b{\mathcal R}&=\nabla_a\nabla_b{\mathcal R}_+
\Theta\left(l\right)+
\nabla_a\nabla_b{\mathcal R}_-\Theta\left(-l\right)+
\nonumber \\
&\epsilon
n_a\left[\partial_b{\mathcal R}\right]\delta\left(l\right)\,,
\label{second2}
\end{eqnarray}
and then,
$\Box {\mathcal R}=\Box {\mathcal R}_+\Theta\left(l\right)+
\Box {\mathcal R}_-\Theta\left(-l\right)+\epsilon
n^a\left[\partial_a{\mathcal R}\right]\delta\left(l\right)$.
Taking the trace of Eq.~\eqref{riccirel} written in terms of the
distribution functions and using the expansions
given in Eqs.~(\ref{second1}) and ~(\ref{second2})
for the second
order derivatives, one obtains
\be\label{derivrel}
f_{\mathcal R R}n^a\left[\partial_aR\right]+f_{\mathcal R\mathcal R}
n^a\left[\partial_a\mathcal R\right]=0\,,
\ee
which is the equation that relates the jumps of the derivatives of $R$
and $\mathcal R$.

Given the second order derivatives of $R$ and $\mathcal R$, in the
distribution formalism, see Eqs.~(\ref{second1}) and ~(\ref{second2}),
the left-hand side of the field equation in Eq.~\eqref{genfield2} thus
depends on terms proportional to the delta function
$\delta\left(l\right)$.  These terms are associated with the presence
of a thin-shell at the separation hypersurface $\Sigma$.  To find the
properties of the thin shell, i.e., the stress-energy tensor for this
hypersurface, let us write the stress-energy tensor in the geometrical
representation $T_{{\rm gr}\hskip 0.03cm ab}$, which we write simply
as $T_{ab}$ to shorten the notation, as a distribution function of the
form
\be\label{tabshellgen}
T_{ab}=T_{ab}^+\Theta\left(l\right)+
T_{ab}^-\Theta\left(-l\right)+\delta\left(l\right)S_{ab},
\ee
where $T_{ab}^+$ is the stress-energy tensor in the geometrical
representation in the region $\mathcal V^+$, $T_{ab}^-$ is the
stress-energy tensor in the geometrical representation in the region
$\mathcal V^-$, and where $S_{ab}$ is the 4-dimensional stress-energy
tensor of the thin shell in the geometrical representation, which can
be written as a 3-dimensional tensor at $\Sigma$ as
\be
S_{ab}=S_{\alpha\beta}e^\alpha_a e^\beta_b.
\ee
With these considerations, the $\delta\left(l\right)$ terms in the
field equations Eq.~\eqref{genfield2} at the hypersurface $\Sigma$ can
be written as
$-\left(f_R+f_\mathcal R\right)\epsilon
\left[K_{\alpha\beta}\right]+
h_{\alpha\beta}\left[\left(f_{RR}-\frac{1}{2}f_{\mathcal R R}
\right)\epsilon n^c\left[\partial_cR\right]+
\left(f_{R\mathcal R}-\frac{1}{2}f_{\mathcal R\mathcal R}\right)
\epsilon n^c\left[\partial_c\mathcal R\right]\right]\break
= 8\pi S_{\alpha\beta}
$,
where $f$ and its derivatives are evaluated considering $R=R_\Sigma$
and $\mathcal R=\mathcal R_\Sigma$. In this step, we used the property
$n_ae^a_\alpha=0$. The jumps of the derivatives of $R$ and $\mathcal
R$ are not independent of each other, and the relationship between
them has been obtained, see Eq.~\eqref{derivrel}.
Inserting Eq.~\eqref{derivrel} into the expression just written for
$S_{\alpha\beta}$
and raising one
index using the inverse induced metric $h^{\alpha\beta}$, yields
finally the equation that allows to compute the stress-energy tensor
of the thin shell as
\be
\epsilon\delta_\alpha^\beta n^c
\left[\partial_cR\right]\left(f_{RR}-
\frac{f_{\mathcal R R}^2}{f_{\mathcal R\mathcal R}}\right)-
\left(f_R+f_\mathcal R\right)\epsilon\left[K_\alpha^\beta\right]
=8\pi S_\alpha^\beta\,,
\ee
where $\delta_\alpha^\beta=h^{\beta\gamma}h_{\gamma\alpha}$ is the
identity matrix.

Several remarks should be done.
Note that the trace of the extrinsic curvature $K$ can be
written in terms of the trace of the extrinsic curvature
$\mathcal K$ and
the conformal factor $f_\mathcal R$ as $\mathcal K = K +
n^c\partial_cf_\mathcal R/f_\mathcal R$. The conformal
factor $f_\mathcal
R\left(R,\mathcal R\right)$ is a function of only $R$ and $\mathcal
R$, and both these variables must be continuous to satisfy their own
junction conditions in Eqs.~\eqref{juncric} and
\eqref{juncriccal}. Consequently, we have $\left[f_\mathcal
R\left(R,\mathcal R\right)\right]=0$. Thus, taking the jump of
$\mathcal K$, expanding $\partial_cf_\mathcal R$ in terms of
$\partial_c
R$ and $\partial_c\mathcal R$, and using Eqs.~\eqref{junction1b} and
\eqref{derivrel}, we recover Eq.~\eqref{junction1b2}. Thus, these two
junction conditions are also not independent.

To conclude, the full set of independent junction conditions of the
theory in the geometrical representation for a
matching with a shell is thus
\begin{eqnarray}\label{fullsetshell}
&\left[h_{\alpha\beta}\right]=0\,, \nonumber
\\
&\left[K\right]=0\,, \nonumber
\\
&\left[R\right]=0\,, 
\\
&\left[\mathcal R\right]=0\,,\nonumber
\\
&\epsilon\delta_\alpha^\beta n^c
\left[\partial_cR\right]\left(f_{RR}-
\frac{f_{\mathcal R R}^2}{f_{\mathcal R\mathcal R}}\right)-
\left(f_R+f_\mathcal R\right)\epsilon\left[K_\alpha^\beta\right]
=8\pi S_\alpha^\beta\,,\nonumber 
\\
&f_{\mathcal R R}n^a\left[\partial_aR\right]+f_{\mathcal R\mathcal R}
n^a\left[\partial_a\mathcal R\right]=0\,, \nonumber
\end{eqnarray}
so composed of six equations. Note that in
the case that $f\left(R,\mathcal R\right)$
reduces to an $f(R)$ theory the fourth and sixth
equations of 
Eq.~\eqref{fullsetshell} are identically zero.

\subsubsection{Matching smoothly at $\Sigma$}

We have obtained the junction conditions for
which two spacetimes, $\mathcal V^+$ and $\mathcal V^-$, can be
matched at a given separation hypersurface $\Sigma$ with the presence
of a thin-shell at $\Sigma$ described by a surface stress-energy
tensor $S_\alpha^\beta$. If $S_\alpha^\beta$ vanishes, the matching
between the two spacetimes is smooth, i.e., without the need for a
thin shell at the separation hypersurface.  A smooth matching between
$\mathcal V^+$ and $\mathcal V^-$ can be achieved by imposing another
set of conditions on the geometrical variables.  Indeed, the presence
of the thin-shell is associated with the term in the stress-energy
tensor proportional to $\delta\left(l\right)$ when written in the
distribution formalism, which is then
reflected in the field equations.

For a smooth matching, i.e., for a matching without a thin-shell, one
must guarantee that the terms proportional to $\delta\left(l\right)$
vanish.  Let us now derive the conditions for such to happen
in the geometrical representation of the theory.

The form of the metric for the whole spacetime as given in
Eq.~(\ref{metricsplit}) is still valid for a smooth matching, so
following the same procedure we have again that $h_{\alpha\beta}$, the
induced metric on $\Sigma$, has no jump,
\be
\label{junction1asmooth}
\left[h_{\alpha\beta}\right]=0.
\ee
Now, the Ricci tensor $R_{ab}$
of the metric $g_{ab}$
can be written in terms of distribution functions as
$
R_{ab}=R_{ab}^+\Theta\left(l\right)+
R_{ab}^-\Theta\left(-l\right)
-\big(\epsilon\,
e_a^\alpha e_b^\beta
\left[K_{\alpha\beta}\right]+n_an_b\left[K\right]\big)
\delta\left(l\right)$, 
where $K_{\alpha\beta}=\nabla_\alpha n_\beta$ is the extrinsic
curvature of $\Sigma$ with $n_\beta=e_\beta^b n_b$, and
$K=K^\alpha_\alpha$ is the trace of $K_{\alpha\beta}$.  But the field
equation Eq.~\eqref{genfield2} has an $R_{ab}$ term and so in general
it would possess a term proportional to $\delta\left(l\right)$ which
cannot be present for a smooth matching.  To avoid the presence of
this term in the field equation Eq.~\eqref{genfield2}, one must impose
that the jump of the extrinsic curvature $K_{ab}=e_a^\alpha e_b^\beta
K_{\alpha\beta}$ must vanish, i.e., one obtains the
following junction
condition,
\be\label{smoothgeom1}
\left[K_{\alpha\beta}\right]=0.
\ee
Since $\left[K_{\alpha\beta}\right]=0$ it implies directly that its
trace vanishes, $\left[K\right]=0$.
Moreover $\left[K_{\alpha\beta}\right]=0$ implies that
$R_{ab}=R_{ab}^+\Theta\left(l\right)+ R_{ab}^-\Theta\left(-l\right)$,
and thus $R=R^+\Theta\left(l\right)+R^-\Theta\left(-l\right)$.  As we
have done previously, computing then the partial derivatives of $R$
one finds $\partial_aR^+\Theta\left(l\right)+
\partial_aR^-\Theta\left(-l\right)+\epsilon \left[R\right]
n_a\delta\left(l\right)$, which when put into
the terms $\partial_af_\mathcal R\partial_bf_\mathcal R$ in the field
equation, Eq.~\eqref{genfield2}, gives rise to terms depending on
products of these derivatives, such as $\partial^cR\partial_cR$, which
are singular terms, or on $\Theta\left(l\right)\delta\left(l\right)$,
which are undefined, and cannot be present in any matching, including
a smooth matching, and so this leads to
\be\label{juncricsmooth}
\left[R\right]=0,
\ee
From Eq.~(\ref{juncricsmooth}) we see that $R$ must be continuous
across the hypersurface $\Sigma$. We then denote the value of $R$ at
$\Sigma$ as $R_\Sigma$.  Also from Eq.~(\ref{juncricsmooth}) one finds
that $\partial_a R$ can be written as $ \partial_a
R=\partial_aR^+\Theta\left(l\right)+
\partial_aR^-\Theta\left(-l\right)$

Turning to the independent connection $\hat\Gamma$, as we have seen
the Palatini Ricci tensor
$\mathcal R_{ab}$ is written in terms of $\hat\Gamma$ and its
derivatives, $\mathcal{R}_{ab}=\partial_c
\hat\Gamma^c_{ab}-\partial_b\hat\Gamma^c_{ac}+
\hat\Gamma^c_{cd}\hat\Gamma^d_{ab}-\hat\Gamma^c_{ad}\hat\Gamma^d_{cb}$.
The Ricci tensor $\mathcal R_{ab}$
can be written in terms of distribution functions as
$
\mathcal R_{ab}=\mathcal R_{ab}^+\Theta\left(l\right)+
\mathcal R_{ab}^-\Theta\left(-l\right)
-\big(\epsilon\,
e_a^\alpha e_b^\beta
\left[\mathcal K_{\alpha\beta}\right]+n_an_b
\left[\mathcal K\right]\big)
\delta\left(l\right)$, 
where $\mathcal K_{\alpha\beta}=\nabla_\alpha n_\beta$ is the extrinsic
curvature of $\Sigma$ with $n_\beta=e_\beta^b n_b$, and
$\mathcal K=\mathcal K^\alpha_\alpha$ is the trace of
$\mathcal K_{\alpha\beta}$.

Let us now look at Eq.~\eqref{genfield2} again. We have already
concluded that the term $R_{ab}$ does not have any terms proportional
to $\delta\left(l\right)$ and also for a smooth matching 
$T_{ab}$ cannot have terms proportional
to $\delta\left(l\right)$. Thus, the the differential terms
on $f_\mathcal{R}$ in Eq.~\eqref{genfield2}
cannot have terms proportional
to $\delta\left(l\right)$.
Then, from Eq.~\eqref{riccirel} it is clear that generically
$\mathcal R_{ab}$ cannot have a term proportional
to $\delta\left(l\right)$. Thus, 
the jump of the extrinsic
curvature $\mathcal K_{ab}=e_a^\alpha e_b^\beta \mathcal
K_{\alpha\beta}$ must vanish, i.e., one obtains the following junction
condition,
\be\label{smoothgeom4}
\left[\mathcal K_{\alpha\beta}\right]=0.
\ee
Since $\left[\mathcal K_{\alpha\beta}\right]=0$ it
implies directly that its
trace vanishes, $\left[\mathcal K\right]=0$.
Moreover $\left[\mathcal K_{\alpha\beta}\right]=0$ implies that
$\mathcal R_{ab}=\mathcal R_{ab}^+\Theta\left(l\right)+ \mathcal
R_{ab}^-\Theta\left(-l\right)$,
and thus ${\mathcal R} =\mathcal R^+\Theta\left(l\right)+\mathcal
R^-\Theta\left(-l\right)$.  As we
have done previously, computing then the partial derivatives of $R$
one finds $\partial_a\mathcal R^+
\Theta\left(l\right)+\partial_a\mathcal R^-
\Theta\left(-l\right)+\epsilon
\left[\mathcal R\right] n_a\delta\left(l\right)$, which when put in
the field equations into the terms $\partial_af_\mathcal \mathcal
R\partial_bf_\mathcal \mathcal R$ in the field equation,
Eq.~\eqref{genfield2}, gives rise to terms depending on products of
these derivatives, such as $\partial^c\mathcal R\partial_c\mathcal R$,
which are singular terms, or on
$\Theta\left(l\right)\delta\left(l\right)$, which are undefined, and
cannot be present in any matching, including a smooth matching, and so
this leads to
\be\label{junccalricsmooth}
\left[\mathcal R\right]=0,
\ee
From Eq.~(\ref{junccalricsmooth}) we see that $\mathcal R$ must be
continuous across the hypersurface $\Sigma$. We then denote the value
of $\mathcal R$ at $\Sigma$ as $\mathcal R_\Sigma$. Also from
Eq.~(\ref{junccalricsmooth}) one finds that $\partial_a \mathcal R$
can be written as $ \partial_a
\mathcal R=\partial_a\mathcal R^+\Theta\left(l\right)+
\partial_a\mathcal R^-\Theta\left(-l\right)$.

Furthermore, due to the presence of the second-order derivative terms
$\nabla_a\nabla_b f_X$ and $\Box f_X$ in Eq.~\eqref{genfield2}, there
will be second-order derivative terms of both $R$ and $\mathcal R$ in
the field equations. These second-order derivatives are
$\nabla_a\nabla_bR=\nabla_a\nabla_bR_+\Theta\left(l\right)+
\nabla_a\nabla_bR_-\Theta\left(-l\right)+
\epsilon n_a\left[\partial_bR\right]\delta\left(l\right)$
and $\Box R=\Box R_+\Theta\left(l\right)+
\Box R_-\Theta\left(-l\right)+
n^a\left[\partial_aR\right]\delta\left(l\right)$, see
Eq.~\eqref{second1} and the one following it,
and the same expressions for 
$\nabla_a\nabla_b{\mathcal R}$
and $\Box \mathcal R$, see
Eq.~\eqref{second2} and the one following it,
and they have
terms proportional to $\delta \left(l\right)$. To avoid the
presence of these terms in order to have a smooth matching,
on has to impose that the jump of the first-order
derivatives of $R$ and $\mathcal R$ vanish, i.e., we obtain
the two following
junction conditions 
\be\label{smoothgeom2}
\left[\partial_aR\right]=0,
\ee
\be\label{smoothgeom3}
\left[\partial_a\mathcal R\right]=0.
\ee

Several remarks should be done.  Note that the extrinsic curvature
$\mathcal K_{\alpha\beta}$ can be written in terms of the extrinsic
curvature $K_{\alpha\beta}$ and the conformal factor $f_\mathcal R$ as
$\mathcal K_{\alpha\beta} = K_{\alpha\beta} - X^c_{ab}n_c$, where
$X^c_{ab}= \frac{1}{2f_\mathcal R}\left(g^c_b\partial_a+
g^c_a\partial_b-g_{ab}\partial^c\right)f_\mathcal R$.  The conformal
factor $f_\mathcal R\left(R,\mathcal R\right)$ is a function of only
$R$ and $\mathcal R$, and both these variables must be continuous to
satisfy their own junction conditions in Eqs.~\eqref{juncricsmooth} and
\eqref{junccalricsmooth}. Consequently, we have $\left[f_\mathcal
R\left(R,\mathcal R\right)\right]=0$. Thus, taking the jump of
$\mathcal K_{\alpha\beta}$, expanding $\partial_cf_\mathcal R$ in
terms of $\partial_c R$ and $\partial_c\mathcal R$, and using
Eqs.~\eqref{junction1asmooth}, \eqref{smoothgeom2} and
\eqref{smoothgeom3}, we recover Eq.~\eqref{smoothgeom1}. Thus, these
two junction conditions are not independent.

To conclude, the full set of independent junction conditions of the
theory in the geometrical representation for a smooth matching
is thus
\begin{eqnarray}
\label{fullsetsmooth}
&\left[h_{\alpha\beta}\right]=0\,, \nonumber
\\
&\left[K_\alpha^\beta\right]=0\,, \nonumber
\\
&\left[R\right]=0\,, 
\\
&\left[\mathcal R\right]=0\,,\nonumber
\\
&\left[\partial_aR\right]=0\,, \nonumber
\\
&\left[\partial_a\mathcal R\right]=0\,,\nonumber 
\end{eqnarray}
so composed of six equations.
Note, that the same set of extra junction conditions can be obtained
from the general set in Eq.~\eqref{fullsetshell} by setting
$S_{\alpha\beta}=0$. In this case, since the factors depending on
derivatives of the function $f$ are generally nonzero, putting the
second equation into the trace of the fifth equation of
Eq.~\eqref{fullsetshell} yields the fifth equation of
Eq.~\eqref{fullsetsmooth}, which on putting it back into the fifth
equation of Eq.~\eqref{fullsetshell} yields the second equation of
Eq.~\eqref{fullsetsmooth}.
Note in addition that in
the case that $f\left(R,\mathcal R\right)$
reduces to an $f(R)$ theory the fourth and sixth
equations of 
Eq.~\eqref{fullsetsmooth} are identically zero.

\subsection{Junction conditions for the
scalar-tensor representation of the theory}\label{scalar}

\subsubsection{Matching with a thin-shell at $\Sigma$}

The nomenclature is the same as previously.  $\mathcal V^+$ and
$\mathcal V^-$ are two regions of a four-dimensional spacetime
$\mathcal V$ separated by a hypersurface $\Sigma$.  The metrics in
each region are $g_{ab}^+$ and $g_{ab}^-$, respectively, the
projection vectors at $\Sigma$ are $e^a_\alpha$ and the normal to
$\Sigma$ is $n_a$.  The distribution functions needed are Heaviside
function and the Dirac function. As before,
$\left[X\right]=X^+|_\Sigma-X^-|_\Sigma$ denotes the jump of $X$
across $\Sigma$.

We now derive the junction conditions for the scalar-tensor
representation of the generalized hybrid metric-Palatini gravity.  We
deal with $g_{ab}$ to start with and only after we deal with the
scalar fields $\varphi$ and $\psi$. Some equations are the same
as in the geometrical representation of the theory, but since they
arise now in the context of the scalar-tensor representation
we also write them here to be complete and self-contained.

Let us start with $g_{ab}$.
First of all, note that in the field equations given by
Eq.~\eqref{genein2}, there is only one term that depends on
derivatives of the metric $g_{ab}$, which is the Einstein's tensor
$G_{ab}$.  Thus, the same reasoning as outlined in
Sec.\ref{sec:geomat} can be followed, i.e., we write the metric in the
distribution formalism as
\be\label{strgab}
g_{ab}=g_{ab}^+\Theta\left(l\right)+g_{ab}^-\Theta\left(-l\right).
\ee
The derivative of $g_{ab}$ becomes $\partial_c g_{ab}=(\partial_c
g_{ab}^+)\Theta(l)+(\partial_c g_{ab}^-)\Theta(-l)+\epsilon
\left[g_{ab}\right]n_c\delta(l)$, where the term proportional to
$\delta\left(l\right)$ is problematic, because the correspondent
Christoffel symbols would have products of the form
$\Theta\left(l\right)\delta\left(l\right)$ which are undefined in the
distribution formalism. Therefore one has to impose
$\left[g_{ab}\right]=0$. Moreover, as $g_{ab}$ induces a metric on
$\Sigma$ which is given $h_{\alpha\beta}=g_{ab}e^a_\alpha e^b_\beta$,
such that from the exterior the induced metric is
$h_{\alpha\beta}^+=g_{ab}^+e^a_\alpha e^b_\beta$ and from the interior
the induced metric is $h_{\alpha\beta}^-=g_{ab}^-e^a_\alpha
e^b_\beta$. Consequently, for $h_{\alpha\beta}$ to give a continuous
metric on $\Sigma$ we must have $h_{\alpha\beta}^+-h_{\alpha\beta}^-=0$,
i.e.,
\be
\label{strjunction1a}
\left[h_{\alpha\beta}\right]=0.
\ee
Again this is the same as the first junction condition in general
relativity and should generally hold in numerous theories of
gravity. The derivative of the metric thus becomes
\be
\partial_c g_{ab}=(\partial_c g_{ab}^+)\Theta(l)+
(\partial_c g_{ab}^-)\Theta(-l)\,.
\ee
The Ricci tensor $R_{ab}$ of the metric $g_{ab}$ written in the
distribution formalism is then
$R_{ab}=R_{ab}^+\Theta\left(l\right)+R_{ab}^-
\Theta\left(-l\right)-\big(\epsilon\,e_a^\alpha
e_b^\beta\left[K_{\alpha\beta}\right]+n_an_b\left[K\right]
\big)\delta\left(l\right)$,
and consequently the Ricci scalar $R$ is
$R=R^+\Theta\left(l\right)+R^-\Theta\left(-l\right)-
2\epsilon\left[K\right]\delta\left(l\right)$,
where $K_{\alpha\beta}=\nabla_\alpha n_\beta$ is the extrinsic
curvature of $\Sigma$ with $n_\beta=e_\beta^b n_b$, and
$K=K^\alpha_\alpha$ is the trace of $K_{\alpha\beta}$, which will be
used further down.

We
now turn to the scalar fields $\varphi$ and $\psi$. We start by
writing the two scalar fields in the distribution formalism in the
usual way as
\be\label{phidist}
\varphi=\varphi^+\Theta\left(l\right)+\varphi^-\Theta\left(-l\right)\,,
\ee
\be\label{psidist}
\psi=\psi^+\Theta\left(l\right)+\psi^-\Theta\left(-l\right)\,.
\ee
The derivatives of the scalar fields in this distribution
representation are of the form
$\partial_a\varphi=\partial_a\varphi^+\Theta\left(l\right)+
\partial_a\varphi^-\Theta\left(-l\right)+\epsilon\delta
\left(l\right)\left[\varphi\right]n_a$,
and
$\partial_a\psi=\partial_a\psi^+\Theta\left(l\right)+
\partial_a\psi^-\Theta\left(-l\right)+\epsilon\delta
\left(l\right)\left[\psi\right]n_a$.
Note that in the field
equations given by Eq.~\eqref{genein2}, there are terms that depend on
products of derivatives of the scalar field, such as
$\partial^c\varphi\partial_c\varphi$
or
$\partial^c\psi\partial_c\psi$. These products
would have terms
depending $\delta\left(l\right)^2$, which are divergent, or terms of
the form $\Theta\left(l\right)\delta\left(l\right)$, which are
undefined. Therefore, to avoid the presence of these terms, one has
to impose the following 
junction conditions for the scalar fields
\be\label{strjc1phi}
\left[\varphi\right]=0,
\ee
\be\label{strjc1psi}
\left[\psi\right]=0,
\ee
i.e., the scalar fields must be continuous across the hypersurface
$\Sigma$. Then, let us define the value of the scalar fields at
$\Sigma$ to be $\varphi_\Sigma$ and $\psi_\Sigma$. Now, using
Eqs.~\eqref{strjc1phi} and~\eqref{strjc1psi} we verify that the terms
$\partial^c\varphi\partial_c\varphi$ and
$\partial_a\varphi\partial_b\varphi$ or
$\partial^c\psi\partial_c\psi$ and $\partial_a\psi\partial_b\psi$
in
Eq.~\eqref{genein2} become regular.

Let us now consider the second-order derivative terms of the scalar
fields, i.e, the terms of the form
$\nabla_a\nabla_b\varphi$
or
$\nabla_a\nabla_b\psi$.
In the
distribution formalism,  for the scalar
field $\varphi$
these terms become
\be\label{dphidist}
\nabla_a\nabla_b\varphi=\nabla_a\nabla_b\varphi_+
\Theta\left(l\right)+\nabla_a\nabla_b\varphi_-\Theta\left(-l\right)+
\epsilon\delta\left(l\right)n_a\left[\partial_b\varphi\right],
\ee
and thus we have $\Box\varphi=\Box\varphi_+\Theta\left(l\right)+
\Box\varphi_-\Theta\left(-l\right)+\epsilon\delta\left(l\right)n^a
\left[\partial_a\varphi\right]$.
Likewise, for the scalar
field $\psi$ we obtain 
\be\label{dpsidist}
\nabla_a\nabla_b\psi=\nabla_a\nabla_b\psi_+\Theta\left(l\right)+
\nabla_a\nabla_b\psi_-\Theta\left(-l\right)+\epsilon\delta
\left(l\right)n_a\left[\partial_b\psi\right],
\ee
and also  $\Box\psi=\Box\psi_+\Theta\left(l\right)+\Box\psi_-
\Theta\left(-l\right)+\epsilon\delta\left(l\right)n^a
\left[\partial_a\psi\right]$.

Now we deal with a thin shell that might appear.
The second-order derivative terms of $\varphi$
and $\psi$ will thus
contribute with terms proportional to $\delta\left(l\right)$ in the
left-hand side of Eq.~\eqref{genein2}. These terms are associated with
the presence of a thin-shell at the separation hypersurface
$\Sigma$. To find the properties of this thin-shell, i.e., to obtain
its stress-energy tensor, we write the stress-energy tensor
 in the scalar-tensor
representation $T_{{\rm sr}\hskip 0.03cm ab}$, which we write simply
as $T_{ab}$ to shorten the notation,
as a distribution function of the
form
\be\label{tabshellgenscalr}
T_{ab}=T_{ab}^+\Theta\left(l\right)+
T_{ab}^-\Theta\left(-l\right)+\delta\left(l\right)S_{ab},
\ee
where $T_{ab}^+$ is the stress-energy tensor in the scalar-tensor
representation in the region $\mathcal V^+$, $T_{ab}^-$ is the
stress-energy tensor in the scalar-tensor representation in the region
$\mathcal V^-$, and where $S_{ab}$ is the 4-dimensional stress-energy
tensor of the thin shell in the scalar-tensor representation, which can
again be written as a 3-dimensional tensor at $\Sigma$ as
\be
S_{ab}=S_{\alpha\beta}e^\alpha_a e^\beta_b.
\ee
Using these considerations, the $\delta\left(l\right)$
factors of the
modified field equations given by Eq.~\eqref{genein2} at the
hypersurface $\Sigma$ can be written as $8\pi
S_{\alpha\beta}=\epsilon
h_{\alpha\beta}n^c\left(\left[\partial_c\varphi\right]-
\left[\partial_c
\psi\right]\right)-\left(\varphi_\Sigma-
\psi_\Sigma\right)\epsilon
\left(\left[K_{\alpha\beta}\right]-\left[K
\right]h_{\alpha\beta}\right)$,
where $K_{\alpha\beta}$ is the extrinsic curvature and
$K=K^\alpha_\alpha$ is its trace and we have used 
$n_ae^a_\alpha=0$.
The terms proportional to
$\delta\left(l\right)$ in the scalar field equations given by
Eqs.~\eqref{genkgi} and~\eqref{genkg}  become
$\epsilon n^a\left[\partial_a\varphi\right]=\frac{8\pi}{3}S$
and
$n^a\left[\partial_a\psi\right]=0$,
respectively.
So, to perform the matching one must impose
the additional junction conditions for the scalar fields
\be\label{strjcphi2}
\epsilon n^a\left[\partial_a\varphi\right]=\frac{8\pi}{3}S\,,
\ee
\be\label{strjcpsi2}
n^a\left[\partial_a\psi\right]=0\,.
\ee
Inserting then
Eqs.~\eqref{strjcphi2} and~\eqref{strjcpsi2} 
into the field
Eq.~\eqref{genein2}, we obtain
$-\epsilon\left(\varphi_\Sigma-\psi_\Sigma\right)
\left(\left[K_{\alpha\beta}\right]-h_{\alpha\beta}
\left[K\right]\right)=
8\pi\left(S_{\alpha\beta}-\frac{1}{3}h_{\alpha\beta}
S\right)
$.
Tracing this result with the inverse induced metric $h^{\alpha\beta}$
cancels out the terms depending on $S_{\alpha\beta}$ and we
obtain
\be\label{zerocurv}
\left[K\right]=0\,.
\ee
i.e., the extrinsic curvature $K_{\alpha\beta}$
does not need to be
continuous at the hypersurface $\Sigma$ but it must at least have a
continuous trace across $\Sigma$.
Inserting Eq.~(\ref{zerocurv}) into the expression for
$S_{\alpha\beta}$, i.e.,
$\epsilon
h_{\alpha\beta}n^c\left(\left[\partial_c\varphi\right]-\left[\partial_c
\psi\right]\right)-\left(\varphi_\Sigma-\psi_\Sigma\right)\epsilon
\left(\left[K_{\alpha\beta}\right]-\left[K
\right]h_{\alpha\beta}\right)
8\pi S_{\alpha\beta}
$,
raising one of the
indices using the inverse induced metric $h^{\alpha\beta}$, and using
Eq.~\eqref{strjcphi2} to cancel the term proportional to $S$, yields
finally the condition to compute the stress-energy tensor of the thin
shell
\be\label{shellfinal}
\epsilon\delta_\alpha^\beta n^a\left[\partial_a\varphi\right]-
\epsilon\left(\varphi_\Sigma-\psi_\Sigma\right)
\left[K_\alpha^\beta\right]=8\pi S_\alpha^\beta\,,
\ee
where $\delta_\alpha^\beta=h^{\alpha\gamma}
h_{\gamma\beta}$ is the identity matrix.
A remark should be done.
Equation~(\ref{shellfinal}) supersedes Eq.~(\ref{strjcphi2}).
Indeed,
putting Eq.~(\ref{zerocurv}) into the trace of Eq.~(\ref{shellfinal})
yields Eq.~(\ref{strjcphi2}). So this latter is not independent of
Eq.~(\ref{shellfinal}).

To conclude, the full set of junction conditions of the generalized
hybrid metric-Palatini gravity in the scalar-tensor representation for
a matching with a thin shell is thus
\beq
&\left[h_{\alpha\beta}\right]=0,\nonumber \\
&\left[K\right]=0,\label{fullsetscalarshell}\nonumber \\
&\left[\varphi\right]=0, \nonumber \\
&\left[\psi\right]=0,           \\
&\epsilon\delta_\alpha^\beta n^a\left[\partial_a\varphi\right]-
\epsilon\left(\varphi_\Sigma-\psi_\Sigma\right)
\left[K_\alpha^\beta\right]=8\pi S_\alpha^\beta, \nonumber\\
&\left[\partial_a\psi\right]=0, \nonumber 
\eeq
so composed of six equations. The scalar-tensor representation of the
theory has thus the same number of
junction conditions as the geometrical
representation, as expected
from the equivalence between the two representations.

\subsubsection{Smooth matching at $\Sigma$}

We have obtained the junction conditions for which two spacetimes,
$\mathcal V^+$ and $\mathcal V^-$, can be matched at a given
separation hypersurface $\Sigma$ with the presence of a thin-shell at
$\Sigma$. We now turn to the case of a smooth matching.

For a smooth matching between $\mathcal V^+$ and $\mathcal
V^-$, i.e., for a matching without a thin-shell, one must guarantee
that the terms proportional to $\delta\left(l\right)$ vanish.  Let us
now derive the conditions for such to happen in the scalar-tensor
representation of the theory.

The metric $g_{ab}$ can still be written in the same for as in
Eq.~\eqref{strgab} for the smooth matching case. Thus, following the
same arguments presented in the previous section, we conclude again
that the induced metric $h_{\alpha\beta}$ on $\Sigma$ must be
continuous, i.e.,
\be\label{strsjc1}
\left[h_{\alpha\beta}\right]=0.
\ee
Now, the Ricci tensor $R_{ab}$ of the metric $g_{ab}$ can be written
in terms of distribution functions as
$R_{ab}=R_{ab}^+\Theta\left(l\right)+
R_{ab}^-\Theta\left(-l\right)-\big(\epsilon\,e_a^\alpha
e_b^\beta
\left[K_{\alpha\beta}\right]+n_an_b\left[K\right]\big)
\delta\left(l\right)$,
where $K_{\alpha\beta}=\nabla_\alpha n_\beta$ is the extrinsic
curvature of $\Sigma$ with $n_\beta=e_\beta^b n_b$,
and $K=K^\alpha_\alpha$ is the trace of $K_{\alpha\beta}$.  But the
field equation Eq.~\eqref{genein2} has an $R_{ab}$ term and so in
general it would possess a term proportional to $\delta\left(l\right)$
which cannot be present for a smooth matching.  To avoid the presence
of this term in the field equation Eq.~\eqref{genein2}, one must
impose that the jump of the extrinsic curvature $K_{ab}=e_a^\alpha
e_b^\beta K_{\alpha\beta}$ must vanish, i.e., one obtains the
following junction condition,
\be\label{strsjc2}
\left[K_{\alpha\beta}\right]=0.
\ee
Since $\left[K_{\alpha\beta}\right]=0$ it implies directly that its
trace vanishes, $\left[K\right]=0$.

Furthermore, the same forms of the scalar fields 
$\varphi$ and $\psi$
written in the distribution formalism in
Eqs.~\eqref{phidist} and~\eqref{psidist} 
are also still valid in the smooth matching
case. Consequently, following the reasoning presented in the previous
section, we derive that both scalar fields must be continuous across
$\Sigma$, i.e.,
\be\label{strsjc1phi}
\left[\varphi\right]=0\,.
\ee
\be\label{strsjc1psi}
\left[\psi\right]=0\,.
\ee

Accordingly, the second-order derivative terms of the scalar fields
$\varphi$ and $\psi$
are given by the same expressions as before,
i.e., as represented in Eqs.~\eqref{dphidist} and~\eqref{dpsidist}.
These derivatives have terms proportional to
$\delta\left(l\right)$, which can not be present in a smooth
matching. Consequently, to avoid the presence of these terms in the
field equation Eq.~\eqref{genein2} and the scalar field equations of
motion in Eqs.~\eqref{genkgi} and~\eqref{genkg}, one must impose that
the partial derivatives of $\varphi$ and $\psi$ are continuous across
the hypersurface $\Sigma$, thus having the two junction conditions
\be\label{strsjcphi2}
\left[\partial_a\varphi\right]=0\,,
\ee
\be
\left[\partial_a\psi\right]=0\,.
\ee

To conclude, the full set of junction conditions of the
theory for a smooth matching in the scalar-tensor representation is
\beq
\label{setstrsmooth} 
&\left[h_{\alpha\beta}\right]=0,\nonumber \\
&\left[K_\alpha^\beta\right]=0, \nonumber \\
&\left[\varphi\right]=0, \\
&\left[\psi\right]=0,    \nonumber              \\
&\left[\partial_a\varphi\right]=0, \nonumber \\
&\left[\partial_a\psi\right]=0, \nonumber 
\eeq
so composed of six equations. Note that the same set of equations
could be obtained from Eqs.~\eqref{fullsetscalarshell} by setting
$S_{\alpha\beta}=0$ and $S=0$. 
In this case, the
second equation together with the 
trace of the 
fifth equation of 
Eq.~\eqref{fullsetscalarshell} enforces
the fifth equation of Eq.~\eqref{setstrsmooth},
which upon
replacement back into the fifth equation of 
Eq.~\eqref{fullsetscalarshell} 
then yields the second equation of Eq.~\eqref{setstrsmooth}, and
one recovers the same set of junction conditions.

\section{Junction conditions for the generalized hybrid
metric-Palatini gravity theory
for static spherically symmetric spacetimes}
\label{calculationsforsphericalthinshells}

\subsection{Geometrical representation}

\subsubsection{Matching with a thin-shell}\label{auxsecgeo}

In the geometrical
representation, let us assume a static spherically symmetric spacetime
with an interior, an exterior and a thin spherical shell
at the junction between the two with an
stress-energy tensor
${S_{\rm gr\,\alpha}}^\beta$.
The first five
matching conditions in the thin shell
case given in Eq.~(\ref{fullsetshell})
still hold, namely, 
$\left[h_{\alpha\beta}\right]=0$,
$\left[K\right]=0$
$\left[R\right]=0$
$\left[\mathcal R\right]=0$, and
$f_{\mathcal R R}n^a\left[\partial_aR\right]+f_{\mathcal R\mathcal R}
n^a\left[\partial_a\mathcal R\right]=0$.
Now, the spacetime is represented by the time $t$,
and the spatial spherical
coordinates, the radial coordinate
$r$, and the angles $\theta$ and $\phi$.
The stress-energy tensor ${S_{\rm gr\,\alpha}}^\beta$,
which we write simply as 
$S_\alpha^\beta$ to shorten the notation, 
at the thin shell of radius $r_\Sigma$, can be written as
a perfect fluid stress-energy tensor, i.e.,
\be
 S_\alpha^\beta=\text{diag}\left(-\sigma,p,p\right)\,,
\label{energymomentumt}
\ee
where $\sigma$ is the surface density of the thin shell and
$p$ is
the 
transverse pressure on the thin shell.
The trace of the extrinsic curvature on a spherical
thin shell is $K=K_\alpha^\alpha
=K_0^0+
K_\theta^\theta
+K_\phi^\phi
$, and since $K_\theta^\theta=K_\phi^\phi$, it can be put as
$K=K_0^0+
2K_\theta^\theta$. 
Now, the second junction condition in Eq.~(\ref{fullsetshell})
is
$\left[K\right]=0$, so that for a spherically symmetric
thin shell it holds that 
$\left[K_0^0\right]=-2\left[K_\theta^\theta\right]$.
Then the last equation of Eq.~(\ref{fullsetshell})
has two components,
corresponding to 
$S_0^0=-\sigma$ and $S_\theta^\theta=p$, so that
it reduces to
\be\label{geos}
\frac{\epsilon}{8\pi}\left[\left(f_R+f_\mathcal R\right)
\left[K_0^0\right]-n^c\left[\partial_cR\right]\left(f_{RR}-
\frac{f_{\mathcal R R}^2}{f_{\mathcal R\mathcal R}}\right)\right]
=\sigma\,,
\ee
\be\label{geop}
\frac{\epsilon}{8\pi}
\left[
\frac{1}{2}\left(f_R+f_\mathcal R\right)\left[K_0^0\right]
+
n^c\left[\partial_cR\right]
\left(f_{RR}-
\frac{f_{\mathcal R R}^2}{f_{\mathcal R\mathcal R}}\right)
\right]=p
\,,
\ee
at
$r_\Sigma$. This is the thin shell equation for a
spherically symmetric matching. The other
equations in Eq.~(\ref{fullsetshell}) have to hold also.

\subsubsection{Smooth matching}

If there is no shell then $\sigma=0$ and $p=0$.  From
Eqs.~\eqref{geos} and~\eqref{geop}, we can see that if
$\left[K_0^0\right]=0$ and $\left[\partial_cR\right]=0$, we recover
$\sigma=0$ and $p=0$. Since $\left[K\right]=0$, we have that
$\left[K_\theta^\theta\right]=\left[K_\phi^\phi\right]=-
\frac{1}{2}\left[K_0^0\right]=0$, and thus we conclude that the jump
of the extrinsic curvature vanishes identically, i.e., $
\left[K_{\alpha\beta}\right]=0 $.  The other equations in
Eq.~(\ref{fullsetsmooth}) have to hold also.

\subsection{Scalar-tensor representation}

\subsubsection{Matching with a thin-shell}\label{auxsecsca}

In the scalar representation, the
stress-energy 
tensor ${S_{{\rm sr}\;\alpha}}^\beta$, which we write
simply as $S_\alpha^\beta$ to shorten the notation,
is a diagonal
matrix which can be written as
\be
S_\alpha^\beta=\text{diag}\left(-\sigma,p,p\right)\,,
\label{energymomentumscalarrep}
\ee
where again
$\sigma$ is the surface density of the thin shell and $p$ is
the 
transverse pressure on the thin shell.
Using this representation and
noticing that, since $\left[K\right]=0$, the angular components of
$\left[K_\alpha^\beta\right]$ are the same and
$\left[K_0^0\right]=-2\left[K_\theta^\theta\right]$, then we can use
Eqs.~\eqref{strjcphi2} 
and 
\eqref{shellfinal}
in the form
$\epsilon n^a\left[\partial_a\varphi\right]=\frac{8\pi}{3}
\left(2p-\sigma\right)$
and
$\frac{16\pi}{3}\left(\sigma+p\right)=\epsilon\left(
\varphi_\Sigma-\psi_\Sigma\right)\left[K_0^0\right]$,
respectively,
to solve the system for $\sigma$ and $p$ as
\be\label{scas}
\frac{\epsilon}{8\pi}\Big[\left(
\varphi_\Sigma-\psi_\Sigma\right)\left[K_0^0\right]-
n^a\left[\partial_a\varphi\right]\Big]=\sigma\,,
\ee
\be\label{scap}
\frac{\epsilon}{8\pi}\left[
\frac{1}{2}\left(\varphi_\Sigma-\psi_\Sigma
\right)\left[K_0^0\right]
+
n^a\left[\partial_a\varphi
\right]
\right]
=p\,,
\ee
at
$r_\Sigma$.
Note that this result is equivalent to the one obtained in the
geometrical representation of the theory.
Indeed, when we map back
$\varphi=f_R$, use Eq.~\eqref{partialf} to expand the partial
derivatives of $f_R$, and use Eq.~\eqref{derivrel} to relate the
partial derivatives of $R$ and $\mathcal R$, we recover
Eqs.~\eqref{geos} and~\eqref{geop}.
The other
equations in Eq.~(\ref{fullsetscalarshell}) have to hold also.

\subsubsection{Smooth matching}

If there is no shell then $\sigma=0$ and $p=0$.
From Eqs.~\eqref{scas} and~\eqref{scap}, we can see that if
$\left[K_0^0\right]=0$ and $\left[\partial_a\varphi\right]=0$,
we recover
$\sigma=0$ and $p=0$. Since $\left[K\right]=0$, we have that
$\left[K_\theta^\theta\right]=\left[K_\phi^\phi\right]=-
\frac{1}{2}\left[K_0^0\right]=0$,
and thus we conclude that the jump of the extrinsic curvature vanishes
identically, i.e.,
$
\left[K_{\alpha\beta}\right]=0
$. 
The other
equations in Eq.~(\ref{setstrsmooth}) have to hold also.

\section{First application: A star. Thin shell:
Matching an interior Minkowski spacetime to an exterior Schwarzschild
spacetime}
\label{app}

\subsection{Geometrical representation}

\subsubsection{The theory and the configuration}

The first step is to choose a theory, i.e., a form for the function
$f\left(R,\mathcal R\right)$. In this case, we chose
\be\label{partf}
f\left(R,\mathcal R\right) =
R+\mathcal R+\frac{R \mathcal R}{R_0}.
\ee
For this particular form of the function $f$, the first and second
derivatives $f_R$, $f_\mathcal{R}$, $f_{RR}$, $f_{\mathcal R\mathcal
R}$ and $f_{R\mathcal R}$ become
$
f_R=1+\bar{\mathcal R}, \quad f_\mathcal R=1+\bar{R}
$,
$
f_{RR}=f_{\mathcal R\mathcal R}=0, \quad f_{R\mathcal R}=\frac{1}{R_0}
$.
See the Appendix~\ref{choosingatheory}
for the rationale of the choice of Eq.~(\ref{partf}).

The configuration for which we want to find a solution is
a star shell.  It is our first application of the use of the junction
conditions for the hybrid metric-Palatini gravity theory derived above
to match two different spacetimes with a thin shell in between.
The 
interior is a Minkowski spacetime
and the exterior is
a Schwarzschild spacetime.

\subsubsection{The interior and the exterior solutions}

Considering the 
interior as the Minkowski spacetime
and the exterior as
the Schwarzschild spacetime
we note that
both these spacetimes are solutions of the modified field
equations in vacuum with $T_{ab}=0$. Both the Minkowski and the
Schwarzschild solutions have a vanishing Ricci tensor
$R_{ab}=0$ and, consequently, a vanishing Ricci scalar
$R=0$. Inserting the form of $f\left(R,\mathcal R\right)$
given in Eq.~\eqref{partf} and its appropriate derivatives
into Eq.~\eqref{genfield2}, one obtains
that $\mathcal R_{ab}=0$, and thus
$\mathcal R=0$. This means that the function $f\left(R,\mathcal
R\right)$ vanishes both in the interior and the exterior solutions.
Thus the interior and exterior spacetimes
are characterized by the 
 Minkowski and the 
Schwarzschild line elements, respectively, and
by a function $f$ given by 
$f\left(R,\mathcal
R\right)=0$.

In the usual spherical coordinates
$\left(t,r,\theta,\phi\right)$ we can write
for the interior
\be\label{mink}
ds_i^2=-dt^2+dr^2+r^2d\Omega^2\,,\quad
f\left(R_i,\mathcal
R_i\right)=0\,,\quad 0\leq r\leq r_\Sigma\,,
\ee
where the subscript $i$ denotes interior,
with  $d\Omega^2=d\theta^2+\sin^2\theta
d\phi^2$ being the line element over
the unit sphere, 
$r_\Sigma$ being the boundary radius between the
interior and the exterior,
and we have not put a subscript $i$
in the coordinates to not overcrowd the notation.
Note that from this point onwards we will use
the subscript $i$ for the interior region instead
of the subscript
- that we have used for the general formalism.

In the same spherical coordinates
we can write
for the exterior
\begin{eqnarray}\label{schw}
ds_e^2=&-\left(1-\frac{2M}{r}\right)\alpha dt^2+
\dfrac{dr^2}{1-\frac{2M}{r}}+r^2d\Omega^2,\nonumber\\
&f\left(R_e,\mathcal
R_e\right)=0\,,\quad  r_\Sigma\leq r<\infty\,,
\end{eqnarray}
where the subscript $e$ denotes exterior, 
$M$ is the Schwarzschild mass,
and $\alpha $ is a dimensionless 
constant that guarantees that the interior and exterior time
coordinates are the same,
and we have not put a subscript $e$
in the coordinates to not overcrowd the notation.
Note that from this point onwards we will use
the subscript $e$ for the interior region instead
of the subscript
$+$ that we have used for the general formalism.

\subsubsection{The matching with a thin shell}

Let us now apply the junction conditions
given in
Eq.~\eqref{fullsetshell}, see also
Eqs.~\eqref{geos} and~\eqref{geop},
to perform a
matching between these two spacetimes.

The first junction condition in Eq.~\eqref{fullsetshell}, 
$\left[h_{\alpha\beta}\right]=0$,
sets the expression for the constant $\alpha$ as
$\alpha=\left(1-\frac{2M}{r_\Sigma}\right)^{-1}$
where $r_\Sigma$ is the radius at which we perform the matching.

The second junction condition of Eq.~(\ref{fullsetshell}),
$\left[K\right]=0$, has to be worked carefully. Using the metrics in
Eqs.~\eqref{mink} and \eqref{schw}, one verifies that for a matching
performed at radius $r_\Sigma$ we have
$\left[K_0^0\right]=\frac{M}{r_\Sigma^2}
\frac1{\sqrt{1-\frac{2M}{r_\Sigma}}}$,
$\left[K_\theta^\theta\right]=
\left[K_\phi^\phi\right]=\frac{1}{r_\Sigma}
\left(\sqrt{1-\frac{2M}{r_\Sigma}}-1\right)$.
The trace of the extrinsic curvature on a spherical
thin shell is $K=K_\alpha^\alpha
=K_0^0+
K_\theta^\theta
+K_\phi^\phi
$, and since $K_\theta^\theta=K_\phi^\phi$, it can be put as
$K=K_0^0+
2K_\theta^\theta$. 
So,
taking the
trace, we obtain
$\left[K\right]=-\frac{2}{r}-\frac{3M-2r}{r^2\sqrt{1-\frac{2M}{r}}}$
and thus this junction condition becomes effectively a constraint on
the radius $r_\Sigma$ at which the matching must be performed,
namely, $-\frac{2}{r}-\frac{3M-2r}{r^2\sqrt{1-\frac{2M}{r}}}=0$.
Solving
this constraint for $r_\Sigma$ yields
\be\label{cond}
r_\Sigma = \frac{9}{4}M.
\ee
Thus $\alpha$ above 
has the value $\alpha=9$. 
Here there is a difference between general relativity and the
generalized hybrid metric-Palatini gravity and it is also a feature of
other $f\left(R\right)$ theories. In general relativity,
the matching between these two spacetimes could be performed for any
value of the radial coordinate as long as  $r_\Sigma>2M$,
i.e., $r_\Sigma$ is larger than the gravitational radius of the
Schwarzschild solution. However, in here we are forced to perform the
matching at a specific value of $r$ to satisfy the extra
$\left[K\right]=0$ junction condition.
Note that $\frac{9}{4}M$ in Eq.~(\ref{cond})
corresponds to the Buchdahl limit for compactness of a fluid
star. In general
relativity, the same limit arises for thin-shells
with surface density $\sigma$
and pressure $p$
if one imposes that the equation of state
for the matter in the thin shell
obeys $2p\leq\sigma$, with the
Buchdahl limit arising when the inequality is saturated.
In the context of thin shells in general relativity
the radius given in Eq.~\eqref{cond}
was first found in \cite{sen}.

The third junction condition of
Eq.~\eqref{fullsetshell} is $\left[R\right]=0$.  Both  metrics in
Eqs.~\eqref{mink} and~\eqref{schw} have an identically vanishing Ricci
tensor, i.e., $R_{ab}=0$. As a consequence, they both have a vanishing
Ricci scalar $R=0$ and thus the condition $\left[R\right]=0$ is
automatically satisfied.

The fourth junction condition of
Eq.\eqref{fullsetshell} is $\left[\cal R\right]=0$, and
as we have seen, for a $f\left(R,\mathcal R\right)$
as given in Eq.~\eqref{partf},
when $R_{ab}=0$ one has $\mathcal R_{ab}=0$, and
consequently $\mathcal R=0$, for both metrics inside and outside.
Thus, the condition $\left[\mathcal R\right]=0$ is also automatically
satisfied.

The fifth
junction condition of  Eq.~(\ref{fullsetshell}),
namely, $\epsilon\delta_\alpha^\beta n^c
\left[\partial_cR\right]\left(f_{RR}-
\frac{f_{\mathcal R R}^2}{f_{\mathcal R\mathcal R}}\right)-
\left(f_R+f_\mathcal R\right)\epsilon\left[K_\alpha^\beta\right]
=8\pi S_\alpha^\beta$ is more elaborated and
we need to dwell on it.  Indeed, notice that the jump in the extrinsic
curvature $\left[K_{ab}\right]$ is not zero, and thus we need a thin
shell at $r=R_\Sigma$ to perform the matching. Let us now study the
stress-energy tensor of this thin shell. Following the analysis from
Sec. \ref{auxsecgeo}, the energy density $\sigma$ and the pressure $p$
of the thin-shell are given by Eqs.~(\ref{geos}) and~(\ref{geop}).
From these two equations
we
obtain for our specific case, i.e.,
for an $f\left(R,\mathcal
R\right)$ as given in Eq.~\eqref{partf}
together with its derivatives,
using $\epsilon=1$ since $n^a$ points in the radial
direction and thus is a spacelike vector, 
and using that
since 
$\left[K_0^0\right]=\frac{M}{r_\Sigma^2}
\frac1{\sqrt{1-\frac{2M}{r_\Sigma}}}$
and $r_\Sigma = \frac{9}{4}M$,
see Eq.~\eqref{cond},
one has 
$\left[K_0^0\right]|_{r_\Sigma}=\frac{16}{27M}$,
the following surface energy density $\sigma$
and surface pressure $p$,
\be\label{specS}
\sigma=\frac{4}{27\pi M}\,,
\ee
\be\label{specSp}
p=\frac{2}{27\pi M}.
\ee
From these equations,
one has
the equation of state $\sigma=2p$.
Since both $\sigma$ and $p$ are positive
and $\sigma=2p$,
all the energy
conditions, namely, the null energy condition
(NEC), the weak energy condition (WEC),
the dominant energy condition (DEC),
and the strong energy condition (SEC)
are satisfied at the shell.
Note also that in general relativity
a shell
with $\sigma=2p$ has to be matched at
the Buchdahl radius
$r_\Sigma = \frac{9}{4}M$.
The difference from the generalized
hybrid metric-Palatini gravity
to general relativity, is that in the
former the matching has to be
done at that radius, whereas in the latter
the matching can be done at any other
radius with $\sigma$ and $p$ having
some other equation of state. 

The sixth junction condition of Eq.~\eqref{fullsetshell},
$f_{\mathcal R R}n^a\left[\partial_aR\right]+f_{\mathcal R\mathcal
R}n^a\left[\partial_a\mathcal R\right]=0$,
allows us to infer that as both
$R$ and $\mathcal R$ are identically zero for both the metrics given
in Eqs.~\eqref{mink} and~\eqref{schw}, then $\partial_a R=0$ and
$\partial_a\mathcal R=0$ throughout. Consequently, this junction
condition is also automatically satisfied for these two metrics.

The full solution for the star thin shell in the generalized hybrid
metric-Palatini gravity in the geometrical representation is thus
determined, it is given by Eqs.~\eqref{mink}-\eqref{specSp}.

\subsection{Scalar-tensor representation}

\subsubsection{The theory and the configuration}

Let us start by choosing a theory. Here, we want
the theory in 
the scalar-tensor representation
that corresponds to
the theory in 
the geometrical representation given above. 
So, using the same
choice of the function $f$ given by Eq.\eqref{partf},
the scalar fields
$\psi$ and $\varphi$ can be written as functions of $R$ and
$\mathcal R$ as
$
\psi=-\left( 1+\frac{R}{R_0}\right)
$,
$
\varphi=1+\frac{\mathcal R}{R_0}
$,
or inverting,
$R=-R_0\left(\psi+1\right)$
and
$\mathcal R=R_0\left(\varphi-1\right)$, respectively.
This invertibility allows us to find the form of the
potential $V\left(\varphi,\psi\right)$ associated with the specific
choice of the function $f$, using the equation
$
V\left(\varphi,\psi\right)=-f\left(\alpha,\beta\right)+
\varphi\alpha-\psi\beta
$
derived above, from which we
obtain that the  scalar-tensor representation
of the theory we want to study is given by
the potential 
\be\label{specpot}
V\left(\varphi,\psi\right)=V_0
\left(\psi+1\right)
\left(\varphi-
1\right),
\ee
where $V_0=-3R_0$ is a constant.

The configuration for which we want to find a solution
in the scalar-tensor representation
is again a star shell
solution.
In this way, the correspondence between the geometrical
and the scalar-tensor representations stands out clearly.
Thus, the spacetime is composed of an 
interior which is  Minkowski, a thin shell, 
and an exterior which is
Schwarzschild.

\subsubsection{The interior and the exterior solutions}

Inserting the metrics for the
Minkowski and Schwarzschild spacetimes
and the potential in
Eq.~\eqref{specpot}, into the modified field equations for the
scalar-tensor representation from Eq.~\eqref{genein2}, yields a
partial differential equation
for the fields $\varphi$ and $\psi$. This equation, along with the two
scalar field equations given by
Eqs.~\eqref{genkg} and~\eqref{genkgi}, is a set of three
equations for the two scalar fields $\psi$ and 
$\varphi$. The solution for these equations is $\varphi=1$
and $\psi=-1$, constants. The same result can be obtained directly from
the equations
$
\varphi=1+\frac{\mathcal R}{R_0}
$
and
$
\psi=-\left( 1+\frac{R}{R_0}\right)
$
by setting $R=0$ and ${\mathcal R}=0$.

Thus the interior line element
in spherical
coordinates $\left(t,r,\theta,\phi\right)$,
and  the interior
scalar fields are given by
\begin{eqnarray}\label{mink1}
&ds_i^2=- dt^2+
{dr^2}
+r^2d\Omega^2\,,&\nonumber\\
&\varphi_i=1,\quad\psi_i=-1\,,
\quad 0\leq r\leq r_\Sigma&\,,
\end{eqnarray}
where the subscript $i$ denotes interior,
with  $d\Omega^2=d\theta^2+\sin^2\theta
d\phi^2$ being the line element over
the unit sphere, 
$r_\Sigma$ being the boundary radius between the
interior and the exterior,
and we have not put a subscript $i$
in the coordinates to not overcrowd the notation.

Likewise, the exterior
line element, and the exterior
scalar fields  are given by
\begin{eqnarray}\label{schw1}
ds_e^2=&-\left(1-\frac{2M}{r}\right)\alpha dt^2+
\dfrac{dr^2}{1-\frac{2M}{r}}
+r^2d\Omega^2,\nonumber\\
&\varphi_e=1,\quad\psi_e=-1\,,
\quad r_\Sigma\leq r<\infty\,,
\end{eqnarray}
where the subscript $e$ denotes exterior, 
$M$ is the Schwarzschild mass,
and $\alpha $ is a dimensionless 
constant that guarantees that the interior and exterior time
coordinates are the same,
and we have not put a subscript $e$
in the coordinates to not overcrowd the notation.

\subsubsection{The matching with a thin shell and the full solution}

Let us now apply the junction conditions
given in
Eq.~\eqref{fullsetscalarshell}, see also
Eqs.~\eqref{scas} and~\eqref{scap},
to perform a
matching between these two spacetimes.

The first junction condition in Eq.~\eqref{fullsetscalarshell}, 
$\left[h_{\alpha\beta}\right]=0$,
sets the expression for the constant $\alpha$ as
$\alpha =\left(1-\frac{2M}{r_\Sigma}\right)^{-1}
$,
where $r_\Sigma$ is the radius at which we perform the matching.

The second junction condition of Eq.~(\ref{fullsetscalarshell}),
$\left[K\right]=0$, has to be worked carefully. Using the metrics in
Eqs.~\eqref{mink} and \eqref{schw}, one verifies that for a matching
performed at radius $r_\Sigma$ we have
$\left[K_0^0\right]=\frac{M}{r_\Sigma^2}
\frac{1}{\sqrt{1-\frac{2M}{r_\Sigma}}}$,
$\left[K_\theta^\theta\right]=\left[K_\phi^\phi\right]=
\frac{1}{r_\Sigma}
\left(\sqrt{1-\frac{2M}{r_\Sigma}}-1\right)$.
The trace of the extrinsic curvature on a spherical
thin shell is $K=K_\alpha^\alpha
=K_0^0+
K_\theta^\theta
+K_\phi^\phi
$, and since $K_\theta^\theta=K_\phi^\phi$, can be put as
$K=K_0^0+
2K_\theta^\theta$. 
So,
taking the
trace, we obtain
$\left[K\right]=-\frac{2}{r}-\frac{3M-2r}{r^2\sqrt{1-\frac{2M}{r}}}$
and thus this junction condition becomes effectively a constraint on
the radius $r_\Sigma$ at which the matching must be performed,
namely, $-\frac{2}{r}-\frac{3M-2r}{r^2\sqrt{1-\frac{2M}{r}}}=0$.
Solving
this constraint for $r_\Sigma$ yields
\be\label{condscalar}
r_\Sigma = \frac{9}{4}M.
\ee
Thus $\alpha$ above has the value
$\alpha=9$.
Again, 
in the
generalized hybrid metric-Palatini gravity 
we are forced to perform the
matching at a specific value of $r$ to satisfy the extra
$\left[K\right]=0$ junction condition, whereas
in general relativity this does not happen.
Again, the radius $\frac{9}{4}M$ in Eq.~(\ref{condscalar})
corresponds to the Buchdahl limit for compactness of a fluid
star in general relativity and
in the context of thin shells in general relativity
this radius
was first found in \cite{sen}.

The third junction condition in Eq.~\eqref{fullsetscalarshell},
$\left[\varphi\right]=0$, is automatically satisfied since the scalar
field $\varphi$ was shown to be constant in both the interior and the
exterior solution, from which we obtain $\varphi_\Sigma=1$.

The fourth junction condition
in Eq.~\eqref{fullsetscalarshell},
$\left[\psi\right]=0$, is 
automatically satisfied,
 since the scalar
field $\psi$ was shown to be constant in both the interior and the
exterior solution, from which we obtain $\psi_\Sigma=1$.

The fifth junction condition in Eq.~\eqref{fullsetscalarshell} is
$\epsilon\delta_\alpha^\beta n^a\left[\partial_a\varphi\right]-
\epsilon\left(\varphi_\Sigma-\psi_\Sigma\right)
\left[K_\alpha^\beta\right]=8\pi S_\alpha^\beta$.
Following now the analysis from Sec. \ref{auxsecsca},
the energy density $\sigma$ and the pressure $p$ of the
thin-shell are given by
Eqs.~(\ref{scas}) and~(\ref{scap}).
Since the scalar fields are constant through the two spacetimes, the
junction conditions
$\left[\psi\right]=0$, 
$\left[\varphi\right]=0$, 
$\left[\partial_c\psi\right]=0$, and 
$\left[\partial_c\varphi\right]=0$ are
automatically verified. Also, since the metrics are the same, then the
condition $\left[K\right]=0$ also yields the constraint
$r=r_\Sigma=\frac{9M}{4}$, and thus
$\left[K_0^0\right]=\frac{16}{27M}$
at this hypersurface. The density and transverse pressure of the thin
shell are then given as
$\frac{1}{8\pi}\left\{\left(\varphi-\psi\right)
\left[K_0^0\right]\right\}=\frac{4}{27\pi M}=\sigma=2p$, i.e.,
\be
\sigma=\frac{4}{27\pi M}\,,
\label{sigmashell2}
\ee
\be
p=\frac{2}{27\pi M}\,.
\label{pshell2}
\ee
This is in agreement with Eqs.~\eqref{specS} and~\eqref{specSp}, as
expected.  All the energy conditions, namely, NEC, WEC, DEC, and SEC
are satisfied at the shell.

The sixth
junction condition in Eq.~\eqref{fullsetscalarshell},
$\left[\partial_a\psi\right]=0$, is also automatically satisfied since
the scalar field $\psi$ is a constant and thus $\partial_a\psi=0$ for
both the interior and exterior solutions.

The full solution for the star thin shell in the generalized hybrid
metric-Palatini gravity in the scalar tensor representation is thus
determined, it is given by Eqs.~\eqref{mink1}-\eqref{sigmashell2}.
It has been made evident that both representations, geometrical and
scalar-tensor, give the same results. This full correspondence
between the representations is expected to hold in principle
always, or at least in those cases that there are no
unexpected singularities in one or more fields
of one of the representations.





%


\section{Second application: A quasistar with a black hole.
Smooth matching and thin shell:
Matching two Schwarzschild spacetimes using a perfect fluid
thick shell}\label{secondapplication}

\subsection{Geometrical representation}

The first step is to choose a theory, i.e., 
a form for the function
$f\left(R,\mathcal R\right)$. In this case, we chose
\be
\label{theoryforsmooth}
f\left(R,\mathcal
R\right)=R \,g\left(
\frac{\mathcal R}{R}
\right)+{\mathcal R} \,h\left(
\frac{\mathcal R}{R}
\right)\,,
\ee
for some well-behaved functions $g$ and $h$.

The solution we are looking for is a spherically-symmetric quasistar, i.e., a central Schwarzschild black-hole surrounded by a thick-shell of matter, which in turn is surrounded by an exterior vacuum spacetime.
In this second application, one can  show that the interior junction between the spacetimes
just described have a thin-shell, whereas the exterior junction is
a smooth one. 

We have seen that the geometrical and the scalar representations of the
generalized hybrid metric-Palatini theory
give the same solution. So we will not do the
matching in the geometrical representation and pass directly
to the scalar tensor representation.

\subsection{Scalar-tensor representation}

\subsubsection{The theory and the configuration}

Let us start by choosing a theory. Here, we want
the theory in 
the scalar-tensor representation
that corresponds to the 
the theory in 
the geometrical representation given above
in Eq.~(\ref{theoryforsmooth}).
The potential $V$ equivalent to Eq.~(\ref{theoryforsmooth})
is using $
V\left(\varphi,\psi\right)=-f\left(\alpha,\beta\right)+
\varphi\alpha-\psi\beta$ derived above, given by
\be
\label{potentialforsmooth}
V\left(\varphi,\psi\right)=0\,.
\ee
Thus, the scalars $\varphi$ ans $\psi$ are massless
and do not interact
between themselves.

The configuration for which we want to find a solution is a quasistar,
i.e., a central Schwarzschild black hole surrounded by matter in a
thick shell, surrounded by vacuum, in a static spherically symmetric
configuration.  It is our second application of the use of the
junction conditions for the hybrid metric-Palatini gravity theory
derived above to match two different spacetimes with a thin shell and
a smooth matching.
So, the spacetime that we study here
consists of an
interior Schwarzschild
black hole spacetime, in the middle, 
a shell of perfect fluid with finite
thickness, and 
an exterior Schwarzschild spacetime.

Let us then assume a general form
of a static
spherically symmetric form for the metric,
a symmetry also inherited  by
the scalar fields $\varphi$ and $\psi$.
So  we write the line element  as
\begin{eqnarray}\label{spher}
ds^2=-e^\zeta dt^2+
\dfrac{dr^2}{1-\frac{2m}{r}}
+r^2d\Omega^2\,,
\end{eqnarray}
where
\be\label{zetaspher}
\zeta=\zeta(r)\,,
\ee
is the redshift function, and 
\be\label{mspher}
m=m(r)\,.
\ee
is the mass function.
We write the scalar fields as
\be\label{varphispher}
\varphi=\varphi(r)\,.
\ee
\be\label{psispher}
\psi=\psi(r)\,,
\ee
We  further assume that the
distribution of matter can be described by an anisotropic 
fluid with stress-energy tensor of the form
$T_a^b=\text{diag}\left(-\rho,p_r,p_t,p_t\right)$,
for some spacetime region in a
range of the radial coordinate $r$, such that
\be\label{Tab}
\rho=\rho\left(r\right)\,,
\ee
\be\label{Tabp}
p_r=p_r\left(r\right)\,,\quad p_t=p_t\left(r\right)\,,
\ee
are the 
energy density, the
radial pressure, and
the transverse pressure, respectively, with 
all being functions of the radial coordinate $r$.

Inserting
Eqs.~\eqref{spher}-\eqref{varphispher} and the ansatz for $T_{ab}$,
Eqs.~\eqref{Tab}-\eqref{Tabp},
into
Eqs.~\eqref{genein2}-\eqref{genkgi} yields a set of five
independent equations which read, after rearrangements,
\beq\label{rho}
&&\frac{2m'}{r^2}\left(\varphi-
\psi\right)-\frac{3}{4}\left(1-\frac{2m}{r}
\right)\frac{\psi'^2}{\psi}+\\
&&+\left(\varphi'-\psi'\right)\left(\frac{3m}{r}+
m'-2\right)-\left(r-2m\right)\left(\varphi''-
\psi''\right)
=8\pi \rho
,\nonumber
\eeq
\beq\label{pr}
&&\frac{\varphi-\psi}{r}
\left[\left(1-\frac{2m}{r}\right)\zeta'-
\frac{2m}{r^2}\right]+\\
&&+\left[\left(\varphi'-\psi'\right)
\left(\frac{2}{r}+\frac{\zeta'}{2}\right)-
\frac{3}{4}\frac{\psi'^2}{\psi}\right]
\left(1-\frac{2m}{r}\right)
=8\pi
p_r
,\nonumber
\eeq
\beq\label{pt}
&&\frac{\varphi-\psi}{2}
\left[\left(1-\frac{2m}{r}\right)
\left(\zeta''+\frac{\zeta'^2}{2}
\right)+\frac{\zeta'}{r}\left(1-m'-
\frac{m}{r}\right)-\right.\nonumber\\
&&-\left.\frac{2}{r^2}\left(m'-
\frac{m}{r}\right)\right]+\left(1-
\frac{2m}{r}\right)\left[\left(
\varphi''-\psi''\right)+\frac{3}{4}
\frac{\psi'^2}{\psi}\right]+\nonumber\\
&&+\left(\varphi'-\psi'\right)\left[
\frac{\zeta'}{2}\left(1-\frac{2m}{r}
\right)+\frac{1-m'}{r}-\frac{m}{r^2}\right]
=8\pi p_t,
\eeq
\begin{eqnarray}
\label{kgphi}
\left(1-\frac{2m}{r}\right)\left(\varphi''+
\frac{\zeta'\varphi'}{2}+
\frac{3\varphi'}{2r}\right)+
\frac{\varphi'}{2r}\left(1-2m'\right)\nonumber\\
=
-\frac{8\pi}{3}\left(\rho-2p_r-p_t\right),
\end{eqnarray}
\be\label{kgpsi}
\psi''-\frac{\psi'^2}{2\psi}+
\frac{\psi'}{2r}\left[4+r\zeta'-
\frac{2\left(m'r-m\right)}{r-2m}\right]=0,
\ee
where a prime denotes a derivative with respect to
$r$.
We now have a set of five independent equations for
seven independent variables, namely,
$\zeta$,
$m$,
$\varphi$,
$\psi$, 
$\rho$, $p_r$, and $p_t$, with all
being functions of the
radius $r$. This means that we have to
impose two independent constraints
to determine the system.

The system we are interested in solving is
static and spherically symmetric and composed of three
regions. The interior region made of a vacuum Schwarzschild
black hole of mass
parameter $m$ and horizon radius $r_h$ up to the inner radius
$r_{\Sigma_i}$ of a thick shell, i.e., a region valid for $r_h\leq r
\leq r_{\Sigma_i}$.  The middle region made of a thick shell with
matter, from the inner radius $r_{\Sigma_i}$ up to its exterior radius
$r_{\Sigma_e}$, i.e., a region valid for $r_{\Sigma_i}\leq r \leq
r_{\Sigma_e}$.  The exterior region made of a vacuum  Schwarzschild
spacetime of mass
parameter $M$, exterior to the thick shell up to infinity, i.e., a
region valid for $r_{\Sigma_e}\leq r \leq \infty$.  In brief, the
system is a black hole surrounded by a finite nonaccreting
matter thick
shell, surrounded by vacuum, it is a quasistar.

\subsubsection{The interior, the middle, and the exterior
solutions}

\noindent
{\small \it The interior solution:
Schwarzschild black hole and constant scalar fields}

\noindent
For the interior region we assume a vacuum solution valid up to an
interior radius $r_{\Sigma_i}$,
or $r_{{}_{\Sigma_i}}$.  The ans\"atze and the equations given
in Eqs.~(\ref{spher})-(\ref{kgphi}) yield a consistent solution,
namely, a black hole solution. 
Since we are interested in the exterior to the black hole,
i.e., exterior to the horizon radius $r_h$,
the
solution is of interest in the region $r_h< r\leq r_{\Sigma_i}$.

The interior
line element is
\begin{eqnarray}\label{mink2}
ds_i^2=-\left(1-\frac{2{m{}}}{r}\right)\alpha{}
dt^2+&
\dfrac{dr^2}{1-\frac{2m}{r}}
+r^2d\Omega^2,
\nonumber\\
&r_h< r\leq r_{\Sigma_i}\,,
\end{eqnarray}
where $m$ is the mass in this region, 
$\alpha{}$ is a quantity that will be determined later,
and
we have not put an index $i$ on the
coordinates to not overcrowd the formulas.
Thus, the metric potentials are
\be\label{zetai}
\zeta_i=\ln\left(1-\frac{2{m{}}}{r}\right)+\ln\alpha\,,\quad
r_h< r\leq r_{\Sigma_i}\,,
\ee
and
\be\label{mi}
m_i=m\,,\quad\quad\quad
r_h< r\leq r_{\Sigma_i}\,,
\ee
with $m$ here a constant, the constant mass parameter.
The black hole horizon radius is given in terms of the mass
by $r_h=2m$.
To  complete the solution for the fields we have
for the scalar fields
\be\label{varphimink2}
\varphi_{i}=\varphi_{i_0}\,,\quad\quad
r_h< r\leq r_{\Sigma_i}\,,
\ee
\be\label{psiimink2}
\psi_i=\psi_{i_0}\,,\quad\quad
r_h< r\leq r_{\Sigma_i}\,.
\ee

Since the interior is a vacuum solution, we have $T_{ab}=0$, and
this can be translated to an energy density
and a pressure as 
\be\label{rhomink2}
\rho_i=0\,,\quad\quad
r_h< r\leq r_{\Sigma_i}\,,
\ee
\be\label{pmink2}
p_{ri}=0\,,\quad p_{ti}=0\,,\quad
r_h< r\leq r_{\Sigma_i}\,.
\ee
The interior solution is thus characterized by
a Schwarzschild black hole, constant scalar fields,
and zero matter fields, i.e., vacuum.

\vskip 0.5cm
\vskip 0.5cm
\noindent{\small \it 
The middle solution: Thick shell}

\noindent
For the middle region we assume a thick shell solution with matter
valid from the interior radius $r_{\Sigma_i}$ up to the exterior
radius $r_{\Sigma_e}$.  The ans\"atze and the equations given in
Eqs.~(\ref{spher})-(\ref{kgphi}) yield a consistent solution, namely,
a physical thick shell.  This thick shell solution thus matches the
vacuum Schwarzschild interior, at $r_{\Sigma_i}$, to an exterior
solution which we assume vacuum Schwarzschild solution at
$r_{\Sigma_e}$.  The range of the coordinate $r$ for the thick shell
region is then $r_{\Sigma_i} \leq r\leq r_{\Sigma_e}$.  We first
present the solution for the thick shell and then show the rationale
employed to obtain it. The constraint we impose is that the energy
density solution is constant, inspired by the general relativistic
Schwarzschild solution with matter. The other constraint that we
impose is an appropriate guess for the mass function.
The solution for this middle region is now presented.

The line element is
\begin{eqnarray}\label{sphermiddle}
ds^2=-
{\rm e}^{\zeta(r)}
dt^2+&
\dfrac{dr^2}{1-\frac{2m(r)}{r}}
+r^2d\Omega^2\,,\\
&r_{\Sigma_i}\leq r \leq r_{\Sigma_e}\nonumber
\,,
\end{eqnarray}
where we use
no subscript
for the variables in this middle
region to distinguish between the
interior $i$  and exterior $e$ regions.
Defining
a supplementary
parameter $\beta$ as
$\beta=\sqrt{\frac{1+\frac{6M}{r_{\Sigma_e}}}
{1-\frac{2M}{r_{\Sigma_e}}}}$,
the solutions for the fields $\zeta(r)$
and 
$m(r)$
of the 
gravitational sector are then
\begin{eqnarray}
\label{zeta}
&\zeta=\zeta_0
+
{(\beta-1)}
\ln\left(\frac{r}{M}\right)\,,
\nonumber\\
&\quad\quad\quad\quad\quad\quad\quad
r_{\Sigma_i}\leq r \leq r_{\Sigma_e},
\end{eqnarray}
i.e., ${\rm e}^{\zeta\left(r\right)}=
{\rm e}^{\zeta_0}\left(\frac{r}{M}\right)^{\beta-1}$,
where 
$\zeta_0$ is an integration constant, 
and
\be\label{mass}
m\left(r\right)=M\,\frac{r}{r_{\Sigma_e}}\,,\quad\quad
r_{\Sigma_i}\leq r \leq r_{\Sigma_e},
\ee
where $M$ is a constant with units of mass that will correspond to the
Schwarzschild spacetime mass of the exterior solution, and
$r_{\Sigma_e}$ is a constant with units of $r$ that, upon matching
with the exterior Schwarzschild spacetime, will correspond to the
radius of the outer surface of the thick shell.
Note that here we have put the mass function as $m(r)$, and the
interior part was characterized by the constant mass $m$. There is no
possibility of confusion, and
it will be seen nonetheless that our use of $m$ for the
interior mass is appropriate.
The solutions for the
scalar field sector  $\varphi(r)$ and
$\psi(r)$  are
\begin{eqnarray}
\label{varphiofr}
\varphi\left(r\right)=&
-\frac{4\pi M^2\rho_0}
{3-7\frac{M}{r_{\Sigma_e}}}\left(\frac{r}{M}\right)^2
+\psi_0+
\varphi_1{\left(\frac{r}{M}\right)}^{-\frac{\beta+1}{2}}\nonumber
\\
&+\varphi_2{\left(\frac{r}{M}\right)}^{\frac{\beta-1}{2}}
,\;\;
r_{\Sigma_i}\leq r \leq r_{\Sigma_e},
\end{eqnarray}
where $\varphi_1$ and $\varphi_2$
are constants of integration, 
and
\be\label{psir}
\psi(r)=\psi_0\,,\quad\quad
r_{\Sigma_i}\leq r \leq r_{\Sigma_e},
\ee
where 
$\psi_0$ is a constant of integration.

The solutions for the matter sector,
namely, $\rho(r)$, $p_r(r)$, and $p_t(r)$ are,
\be\label{rho0}
\rho=\rho_0\,,\quad\quad r_{\Sigma_i}\leq r \leq r_{\Sigma_e},
\ee
\be\label{profr}
p_r=p_r(r)\,,\quad p_t=p_t(r)\,,\quad\quad
r_{\Sigma_i}\leq r \leq r_{\Sigma_e},
\ee
where $\rho_0$ is a constant, the constraint
providing 
the constant density solution,
$\varphi_1$ and $\varphi_2$ are
the previous mentioned constants of integration,
$\psi_0$ is also an imposed constant,
and $p_r(r)$ and $p_t(r)$
are
long expressions, so we do not
write them explicitly here. Instead, we put them in the 
Appendix~\ref{calculationsappendixb}.

The thick shell solution, i.e., the solution for the middle region, is
visualized in the Appendix~\ref{calculationsappendixb}, where all the
fields are plotted.

The rationale to find the solutions just given 
is the following.
We have a set of five independent equations for
seven independent variables, namely,
$\zeta$,
$m$,
$\varphi$,
$\psi$, 
$\rho$, $p_r$, and $p_t$, all
functions $r$. To solve this system we have to
impose two independent constraints.
We assume a Schwarzschild type
interior solution
within the thick shell, i.e.,
we want to find solutions with constant density $\rho_0$,
$\rho=\rho_0$, which is Eq.~(\ref{rho0}).
This sets the right-hand side of Eq.~(\ref{rho0}).
Then we still have a set of five independent equations for
six independent variables, namely,
$\zeta$,
$m$,
$\varphi$,
$\psi$, 
$p_r$, and $p_t$. This means that we still can
impose one further constraints
to determine the system.
The second constraint we choose to impose is on the
function $m(r)$ of the form
$m\left(r\right)=M\,\frac{r}{r_{\Sigma_e}}$,
which is Eq.~(\ref{mass}), where $M$ is a constant with units of mass
that will correspond to the Schwarzschild spacetime mass of the
exterior solution, and $r_{\Sigma_e}$ is
the radius of the outer
surface of the thick shell to be matched to the
exterior Schwarzschild spacetime.
We could have given, more generically, a
mass function of the form
$m\left(r\right)=
M\left(\frac{r}{r_{\Sigma_e}}\right)^{\hskip-0.07cm n}$, with
$n$ an exponent. The choice $n=1$ allows one to find
analytical solutions. Another interesting solution could be analyzed
by setting $n=3$ and then solving the equations numerically. In
any case for $n>0$ it can be seen that for $r=0$ we obtain $m=0$,
i.e., there is no mass at zero radius, so there is no
singularity associated to it, and also that at $r_{\Sigma_e}$ one has
$m\left(r=r_{\Sigma_e}\right)=M$, which is in agreement with the
value for the mass
of the exterior Schwarzschild spacetime. At this point, notice that
Eq.~(\ref{psir}) features a very simple solution given by
$\psi(r)=\psi_0$, where $\psi_0$ is a constant. This is just a
particular solution for this equation, but we shall consider it to
maintain the simplicity of the full solution.
Now, an equation for $\zeta''$ that can be derived from
the system
of equations given in
Eqs.~(\ref{spher})-(\ref{kgphi}) is 
$\left(\zeta''+\frac{\zeta'^2}{2}
\right)\left(1-\frac{2m}{r}\right)-
\frac{4m'}{r^2}-\frac{\zeta'}{r}
\left(\frac{3m}{r}+m'-2\right)=0$, which is
obtained by subtracting Eq.~\eqref{genkg}
from Eq.~\eqref{genkgi}, using Eqs.~\eqref{rho} to~\eqref{pt} to
cancel the terms depending on $\rho$, $p_r$ and $p_t$, and assuming
$\varphi\neq\psi$.
Inserting now Eqs.~\eqref{mass} and~\eqref{psir}
into the equation for $\zeta''$ just derived yields
the following solution for $\zeta(r)$,
$
{\rm e}^{\zeta\left(r\right)}={\rm e}^{\zeta_0}\left(\frac{r}{M}\right)^{\beta-1}
$,
which is Eq.~(\ref{zeta}), 
$\beta$ has been defined before
that equation, and $\zeta_0$ is an integration  constant.
This can be seen as follows. 
Indeed, 
the equation for $\zeta''$
is
a second
order ordinary differential equation for the function
$\zeta\left(r\right)$ which
with the help of Eqs.~\eqref{mass} and~\eqref{psir}
yields the solution,
$\zeta\left(r\right)=\zeta_0-\left(1+\beta\right)
\log\left(\frac{r}{M}\right)+2\log\left[\left(
\frac{r}{M}\right)^\beta+\zeta_1\right]$,
with $\zeta_0$ and $\zeta_1$ being
constants of integration. Leaving $\zeta_0$ as a free parameter
and setting, as an assumption,
$\zeta_1=0$, yields 
Eq.~\eqref{zeta}.
In order to have constant density $\rho=\rho_0$, we insert
Eqs.~\eqref{zeta},~\eqref{mass}, and~\eqref{psir},
into
Eq.~\eqref{rho}, and search for solutions for $\varphi$ that guarantee
that the left hand side of it, and consequently $\rho$, is
constant. The solution for $\varphi$ is
$\varphi\left(r\right)=
\frac{4\pi r^2r_{\Sigma_e}\rho_0}{7M-3r_{\Sigma_e}}+\psi_0+
\varphi_1{r}^{-\frac{\beta+1}2}
+\varphi_2{r}^{\frac{\beta-1}2}
$, 
which is Eq.~(\ref{varphiofr}) after rearrangements. 
Inserting the solutions for $m$, $\zeta$, $\varphi$ and $\psi$ into
Eqs.~\eqref{pr} and~\eqref{pt} yields the results for both $p_r$ and
$p_t$,
$p_r=p_r(r)$,
$p_t=p_t(r)$, which are in  Eq.~(\ref{profr}),
and 
are
long expressions, so we do not
write them explicitly here. Instead, we put them in
Appendix~\ref{calculationsappendixb}.

\vskip 0.5cm

\noindent{\small \it The exterior solution: Schwarzschild spacetime
and constant scalar fields}

\noindent
For the exterior region we assume a vacuum solution valid from
an exterior radius $r_{\Sigma_e}$
up to infinity.
Again,
the ans\"atze and the equations given
in Eqs.~(\ref{spher})-(\ref{kgphi}) yield a consistent solution,
namely, an exterior Schwarzschild solution. 

The exterior
solution line element is
\begin{eqnarray}\label{schw2}
ds_e^2=-\left(1-\frac{2{M{}}}{r}\right)dt^2
+&
\dfrac{dr^2}{1-\frac{2M}{r}}
+r^2d\Omega^2,
\nonumber\\
& r_{\Sigma_e}\leq r<\infty\,,
\end{eqnarray}
where $M$
represents the spacetime mass,
and we have not put an index $e$ on the coordinates
to not overcrowd the formulas.
Thus,
\be\label{zetae}
\zeta_e=\ln\left(1-\frac{2{M{}}}{r}\right)\,,\quad
r_{\Sigma_e}\leq r<\infty,
\ee
and
\be\label{meM}
m_e=M\,,\quad\quad
r_{\Sigma_e}\leq r<\infty,
\ee
with $M$ here a constant, the constant spacetime mass.
The gravitational radius is $r_g=2M$, and there is no
horizon in the exterior.
To  complete the solution we have
for the scalar fields
\be\label{varphischw2}
\varphi_e=\varphi_{e_0}\,,\quad
r_{\Sigma_e}\leq r<\infty\,,
\ee
\be\label{psischw2}
\psi_i=\psi_{e_0}\,,\quad
r_{\Sigma_e}\leq r<\infty\,,
\ee
i.e., they are constants.

Since the exterior is a vacuum solution, we have $T_{ab}=0$, and
this can be translated to an energy density
and a pressure as 
\be\label{rhoschwarz2}
\rho_e=0\,,\quad
r_{\Sigma_e}\leq r<\infty\,,
\ee
\be\label{pschwarz2}
p_{re}=0\,,\quad p_{te}=0\,,\quad
r_{\Sigma_e}\leq r<\infty\,.
\ee
The exterior solution is thus characterized by
a Schwarzschild spacetime, constant scalar fields,
and zero matter fields, i.e., vacuum. There is
no black hole in this region.

Our solution  is now fully characterized and
complete in each region.  We now have to do the
matching of the thick shell solution to the interior and the exterior
solutions to have the full solution.

\subsubsection{The matching of the thick
shell solution and the full solution}

\vskip 0.5cm
\noindent
{\small \it Interior matching of
a Schwarzschild black hole spacetime to a thick shell:}

\vskip 0.1cm
\noindent
We now do the matching between the interior Schwarzschild black hole
spacetime and the thick shell.  The thick shell solution has
divergences for small values of the radial coordinate $r$. Indeed, in
the limit $r\to 0$ the redshift function $\zeta(r)$ diverges and both
pressures $p_r(r)$ and $p_t(r)$ also diverge. In addition, in some
regions, not in the neighborhood of the center, the energy conditions
are violated. Therefore, if we want to have a nonsingular thick shell
with the matter obeying the energy conditions, we have to perform a
matching with some interior solution that we have chosen to be the
Schwarzschild black hole spacetime together with constant fields
$\varphi$ and $\psi$, see Eqs.~\eqref{mink2}-\eqref{varphimink2}.  In
this case, an almost smooth matching between these two spacetimes is
possible, but in fact there will be a shell that appears due to the
scalar fields. So, since, albeit mild, there is a thin shell, one must
use the six conditions in Eq.~(\ref{fullsetscalarshell}).

The interior region is that
of a black hole of mass $m$ and horizon radius
given by
\be
\label{rh=2m}
r_h=2m\,,
\ee
up to $r_{\Sigma_i}$ where the thick shell
starts.
To find a suitable radius $r_{\Sigma_i}$
to perform the matching, let us consider the
validity of the energy conditions. By assumption
the energy density
$\rho=\rho_0$ is positive.
It is found by inspection that for
the thick shell solution for
$r\gtrsim 2.09M$, besides $\rho>0$,
one has
$\rho+p_r>0$, $\rho+p_t>0$, $\rho>|p_r|$, $\rho>|p_t|$, and
$\rho+p_r+2p_t>0$,
and so all the energy conditions, i.e., NEC, WEC, DEC, and SEC, are
satisfied in this region. Also, in this region we note that $p_r>0$
and $p_t>0$, so there are no tensions.  We thus perform the matching
for some $r=r_{\Sigma_i}$ such that
\be
2.09M\leq r_{\Sigma_i} <3M\,,
\label{rangeRi}
\ee
where the number in the
first inequality is a number
approximated to the second decimal place,
and 
the latter inequality comes from the fact that
$r_{\Sigma_i}$ is always less than $r_{\Sigma_e}$
and the fact that $r_{\Sigma_e}=3M$, where $M$
is the mass of the exterior solution, as we will see.

Now, the continuity of the metric $h_{ab}$ has to be imposed, see the
first  equation in Eq.~(\ref{fullsetscalarshell}).  For the
continuity of the metric $h_{ab}$, one has
to use the metrics in
Eqs.~\eqref{spher}  and~\eqref{mink2} 
to conclude that the factor $\alpha{}$
that appears in the interior metric Eq.~\eqref{mink2} 
must have the value 
$\alpha{}=e^{\zeta(r_{\Sigma_i})}\left(1-\frac{2
{m{}}}{r_{\Sigma_i}}\right)^{-1}$.

Now, the continuity of the trace of the
extrinsic curvature $K$ at $r_{\Sigma_i}$ has to be imposed, see the
second equation in Eq.~(\ref{fullsetscalarshell})
at the boundary surface, i.e.,
at the shell, at $r_{\Sigma_i}$.
The components of the 
extrinsic curvature
$K_{i\,ab}(r)$
for the interior spacetime
at $r_{\Sigma_i}$ are 
$K_{i\,00}=-\frac{m}{r^2}
{\left(1-\frac{2m}{r}\right)^{-\frac12}}$,
i.e., at $r_{\Sigma_i}$,
$K_{i\,00}=-\frac{m}{r_{\Sigma_i}^2}
{\left(1-\frac{2m}{r_{\Sigma_i}}\right)^{-\frac12}}$,
$K_{i\,\theta\theta}=
r\sqrt{1-\frac{2{m{}}}{r}}$, i.e., at $r_{\Sigma_i}$,
$K_{i\,\theta\theta}=r_{\Sigma_i}\sqrt{1-
\frac{2{m{}}}{r_{\Sigma_i}}}$, and 
$K_{i\,\phi\phi}= r\sqrt{1-\frac{2{m{}}}{r}}\sin^2\theta$,
i.e., at $r_{\Sigma_i}$,
$K_{i\,\phi\phi}=r_{\Sigma_i}\sqrt{1-
\frac{2{m{}}}{r_{\Sigma_i}}}\sin^2\theta $.
Therefore 
at $r_{\Sigma_i}$ the trace from the interior side is
$K^i=
\frac{2}{r_{\Sigma_i}}
\left(1-\frac{3m}{2r_{\Sigma_i}}\right)
{\left(1-\frac{2{m{}}}{r_{\Sigma_i}}\right)^{-\frac12}}$.
The components of the extrinsic curvature
$K_{ab}(r)$
for the thick shell are
$K_{00}=-
\frac{\beta-1}{2r}\sqrt{1-\frac{2M}{r_{\Sigma_e}}}$,
i.e., for $r_{\Sigma_e}= 3M$ and at $r_{\Sigma_i}$,
$K_{00}=\sqrt\frac{1}{3}\frac1{r_{\Sigma_i}}$,
$K_{\theta\theta}=r\sqrt{1-\frac{2M}{r_{\Sigma_e}}}$,
i.e., for $r_{\Sigma_e}= 3M$ and at $r_{\Sigma_i}$,
$K_{\theta\theta}=\sqrt\frac{1}{3}r_{\Sigma_i}$,
and 
$K_{\phi\phi}=
r\sqrt{1-\frac{2M}{r_{\Sigma_e}}}\sin^2\theta$,
i.e., for $r_{\Sigma_e}= 3M$
and at $r_{\Sigma_i}$,
$K_{\phi\phi}=\sqrt\frac{1}{3}r_{\Sigma_i}\sin^2\theta$.
Therefore 
at $r_{\Sigma_i}$ the trace from the thick shell side is
$K=\frac{\sqrt{3}}{r_{\Sigma_i}}$.
So, 
$[K]=0$ implies 
that
the matching must be performed at a radius
$r_{\Sigma_i}$ given by
\be
r_{\Sigma_i}=3{m{}}\,.
\label{sigmai}
\ee
One can tune the value of ${m{}}$ to select
the radius at which the matching should be performed.
See also that Eq.~(\ref{sigmai}) implies then that the
mass function as given in 
Eq.~(\ref{mass}) is continuous, and
to denote the constant interior mass
by $m$ was a good choice.

The field 
$\varphi_i$ is a constant $\varphi_{0_i}$,
see Eq.~\eqref{varphimink2},
so at $r_{\Sigma_i}$ one must have 
$\varphi_{0_i}=\psi_0+\left(\frac{54M^4}{r_{\Sigma_i}^{2}}+
40Mr_{\Sigma_i}-6r_{\Sigma_i}^{2}\right)\pi\rho_0$,
as it will be clear when we do the matching to
the exterior region.
In this way the 
third condition in
Eq.~(\ref{fullsetscalarshell}), $\left[\varphi\right]=0$,
is satisfied. So, 
\begin{eqnarray}
\label{varphiinteriorreal}
\hskip -0.7cm
\varphi_i=
\psi_0+&
\left[
54
\left(\frac{M}{r_{\Sigma_i}}\right)^2+
40\frac{M}{r_{\Sigma_i}}-6\left(
\frac{r_{\Sigma_i}}{M}\right)^2\right]\pi M^2\rho_0\,,
\\
&\,\quad
r_h\leq r \leq r_{\Sigma_i},
\nonumber
\end{eqnarray}
throughout the interior region.

The field 
$\psi_i$ is a constant $\psi_{0_i}$, see Eq.~\eqref{psiimink2},
so at $r_{\Sigma_i}$, one must have
$\psi_{0_i}=\psi_0$,
where we have used 
Eq.~\eqref{psir}.
In this way the 
fourth condition in
Eq.~(\ref{fullsetscalarshell}), $\left[\psi\right]=0$,
is satisfied. So,
\be
\psi_i=\psi_0\,,\quad\quad r_h\leq r\leq r_{\Sigma_i},
\label{psithinshell}
\ee
throughout the interior
region, and as a matter of fact throughout the thick shell
and the exterior.
The value $\psi_0$ 
is a free
parameter.

We can now see whether there is a thin shell 
or not by computing the components
of the stress-energy tensor $S_\alpha^\beta$ at $r_{\Sigma_i}$.
Using the fifth equation in
Eq.~(\ref{fullsetscalarshell}),
together with the fact that the shell is assumed to be a perfect 
fluid such that the stress-energy tensor is 
$S_\alpha^\beta=\text{diag}\left(-\sigma,p,p\right)$,
where $\sigma$ is the energy density and $p$
the tangential pressure at the shell
one finds,
$\sigma=\frac{\epsilon}{8\pi}
\left(
-
n^a\left[\partial_a\varphi\right]
+\left(
\varphi-
\psi\right)\left[K_0^0\right]
\right)$
and 
$p=\frac{\epsilon}{8\pi}\left(
n^a\left[\partial_a\varphi
\right]+\frac{1}{2}\left(
\varphi-
\psi\right)\left[K_0^0\right]\right)$,
all quantities evaluated at $r_{\Sigma_i}$,
see also Eqs.~(\ref{scas}) and~(\ref{scap}).
One  considers the case 
$\epsilon=1$ since $n^a$
points in the radial direction and thus is a spacelike vector.
Finally, using 
$\left[\partial_c\varphi\right]=-
\partial_r\varphi|_{r_{\Sigma_i}}$, where
$\partial_r\varphi$ is the derivative of
Eq.~\eqref{varphiofr}
with the constants $\varphi_1$ and 
$\varphi_2$ correctly evaluated at the
external boundary, see below, 
and
$\left[K_0^0\right]=\frac{1}{r}
\left(\frac{\frac{{m{}}}{r}}{\sqrt{1-\frac{2{m{}}}{r}}}-
\sqrt{1-\frac{2M}{r_{\Sigma_e}}}\right)
$ evaluated at $r_{\Sigma_i}$, i.e.,
$\left[K_0^0\right]=\frac{1}{r_{\Sigma_i}}
\left(\frac{\frac{{m{}}}{r_{\Sigma_i}}}{\sqrt{1-
\frac{2{m{}}}{r_{\Sigma_i}}}}-
\sqrt{1-\frac{2M}{r_{\Sigma_e}}}\right)
$, i.e.,
$\left[K_0^0\right]=0$ since
$r_{\Sigma_i}=3{m{}}$,
we obtain from
Eqs.~(\ref{scas}) and~(\ref{scap})
for the energy density $\sigma$ and the
pressure $p$ for the thin shell at $r_{\Sigma_i}$ the results
\beq
\hskip -0.9cm
\dfrac{\sigma}{M}=\hskip 0.3cm
\dfrac1{2}
\left(10-\frac{27M^3}{r_{\Sigma_i}^3}-
\frac{3r_{\Sigma_i}}{M}\right)\rho_0,\,\,
r=r_{\Sigma_i},
\label{sigmashell}
\eeq
\beq
\hskip -0.85cm
\dfrac{p}{M}=
-\dfrac1{2}
\left(10-\frac{27M^3}{r_{\Sigma_i}^3}-
\frac{3r_{\Sigma_i}}{M}\right)\rho_0,\,\,
r=r_{\Sigma_i}.
\label{pshell}
\eeq
So at $r_{\Sigma_i}$, for the thin shell, one
has $p=-\sigma$.  Since
$\rho_0>0$, one can verify that in the region $2.09M<r<3M$,
see Eq.~\eqref{rangeRi}, $\sigma>0$, $\sigma+p=0$, and
$\sigma=|p|$, from which we conclude that the NEC, WEC, and DEC are
verified no matter the value of $r_{\Sigma_i}$ that we choose within
this region.  Note that this shell does not contribute to the mass.
This was 
expected, since we have found that the mass $m$ is continuous at
$r_{\Sigma_i}$.

From
Eq.~(\ref{psithinshell}), the junction condition
$\left[\partial_c\psi\right]=0$, see the
sixth condition in Eq.~(\ref{fullsetscalarshell}), is 
trivially satisfied.

\vskip 0.5cm
\noindent
{\small \it Exterior smooth matching of the thick shell to a
Schwarzschild spacetime:}

\vskip 0.1cm
\noindent
We now do the matching between the thick shell and the
exterior Schwarzschild spacetime. In this case, a smooth matching
between these two spacetimes is possible, so
one uses the six conditions in
Eq.~(\ref{setstrsmooth}).
As we will see, the matching to be smooth has to be done necessarily
at
\be
r_{\Sigma_e}=3M\,,
\label{fracmRe}
\ee
i.e., at the exterior photon orbit 
sphere.

First the continuity of the metric $h_{ab}$
at $r_{\Sigma_e}$ has to be imposed,
see the first
equation in Eq.~(\ref{setstrsmooth}).
For the continuity of the metric $h_{ab}$,
note that the quantity 
that appears in 
Eq.~\eqref{zeta} as an exponent has the value
$\beta=3$
for $\frac{M}{r_{\Sigma_e}}=\frac13$, see Eq.~(\ref{fracmRe}).
Then
the component
within the shell
$e^{\zeta}$ that appears in the metric
Eq.~\eqref{spher}, see also Eq.~\eqref{zeta}, 
is now given by 
$e^{\zeta}=e^{\zeta_0}\left(
\frac{r}{M}\right)^{2}$.
The exterior time-time
component of the metric given in Eq.~\eqref{schw2}
is 
$1-\frac{2M}{r}$.
So imposing continuity of the metric 
at $r_{\Sigma_e}$
yields
$e^{\zeta_0}\left(
\frac{r_{\Sigma_e}}{M}\right)^{2}=
1-\frac{2M}{r_{\Sigma_e}}$.
Since $r_{\Sigma_e}=3M$, one has 
$e^{\zeta_0}=\frac1{27}$.
Thus, $e^{\zeta}=\frac1{27}\left(
\frac{r}{M}\right)^{2}$, i.e.,
\be
\zeta(r)=-\ln27
+2\ln\left(
\frac{r}{M}\right)\,,\quad
r_{\Sigma_i}\leq r \leq r_{\Sigma_e}\,.
\label{zetarfinal}
\ee
Then, at 
$r_{\Sigma_i}$ it follows that 
$\zeta(r_{\Sigma_i})=-\ln27
+2\ln\left(
\frac{r_{\Sigma_i}}{M}\right)$,
an expression necessary to determine
the quantity $\alpha$ that appears
in the junction of the interior
region 
with the inner border of the thick shell,
see above.
As well, from Eq.~\eqref{mass}
we have 
\be
m(r)=\frac{r}{3}\,,\quad
r_{\Sigma_i}\leq r \leq r_{\Sigma_e}\,.
\label{massfinalfinal}
\ee

Then, the continuity of 
the extrinsic curvature $K_{ab}$
at $r_{\Sigma_e}$ has to be imposed,
see the second
equation in Eq.~(\ref{setstrsmooth}),
at the boundary, i.e., at $r_{\Sigma_e}$.
The extrinsic curvature
$K_{ab}(r)$
for the thick shell 
is
$K_{00}=-
\frac{\beta-1}{2r}\sqrt{1-\frac{2M}{r_{\Sigma_e}}}$,
i.e., for $r_{\Sigma_e}= 3M$,
$K_{00}=\sqrt\frac{1}{3}\frac1{r_{\Sigma_e}}$,
$K_{\theta\theta}=r\sqrt{1-\frac{2M}{r_{\Sigma_e}}}$,
i.e., for $r_{\Sigma_e}= 3M$,
$K_{\theta\theta}=\sqrt\frac{1}{3}r_{\Sigma_e}$,
and 
$K_{\phi\phi}=
r\sqrt{1-\frac{2M}{r_{\Sigma_e}}}\sin^2\theta$, i.e., for $r_{\Sigma_e}= 3M$,
$K_{\phi\phi}=\sqrt\frac{1}{3}r_{\Sigma_e}\sin^2\theta$.
The extrinsic curvature
$K_{e\hskip0.01cm ab}(r)$
for the exterior spacetime
is
$K_{e\hskip0.01cm 00}=-
\frac{M}{r^2}\sqrt{\left(1-\frac{2M}{r}\right)^{-1}}$,
 i.e., for $r_{\Sigma_e}= 3M$,
$K_{e\hskip0.01cm 00}=\sqrt\frac{1}{3}\frac1{r_{\Sigma_e}}$,
$K_{e\hskip0.01cm \theta\theta}=
r\sqrt{1-\frac{2M}{r}}$, i.e., for $r_{\Sigma_e}= 3M$,
$K_{e\hskip0.01cm
\theta\theta}=\sqrt\frac{1}{3}r_{\Sigma_e}$, and 
$K_{e\hskip0.01cm \phi\phi}=
r\sqrt{1-\frac{2M}{r}}\sin^2\theta$, i.e., for $r_{\Sigma_e}= 3M$,
$K_{e\hskip0.01cm \phi\phi}=
\sqrt\frac{1}{3}r_{\Sigma_e}\sin^2\theta$.
Therefore 
at $r_{\Sigma_e}=3M$ one has
$K_{ab}
\left(r_{\Sigma_e}\right)=K_{e\hskip0.01cm ab}$,
so that $[K_{ab}]=0$, as it should for a
smooth matching. Of course, the solution 
$r_{\Sigma_e}=3M$ is the only possible solution, justifying
thus 
Eq.~\eqref{fracmRe}.

Now we deal with the scalar field $\varphi$. The exterior scalar field
$\varphi$ is a constant $\varphi_{e_0}$, see Eq.~\eqref{varphischw2}.
Within the thick shell, $\varphi$ is a varying function of $r$,
$\varphi(r)$, see Eq.~(\ref{varphiofr}), so in order that the junction
condition $\left[\varphi\right]=0$ is satisfied, see the third
condition in Eq.~(\ref{setstrsmooth}), we have to impose that at the
exterior boundary $r_{\Sigma_e}$ one has
$\varphi_{e_0}=\varphi\left(r_{\Sigma_e}\right)$.
As shown below we find
\begin{eqnarray}
\label{shellphi2}
\hskip -0.8cm
\varphi\left(r\right)&=
\psi_0+
\left[
54
\left(\frac{M}{r}\right)^2+
40\frac{M}{r}-6\left(
\frac{r}{M}\right)^2\right]\pi M^2\rho_0\,,
\\
&\,\quad
r_{\Sigma_i}\leq r \leq r_{\Sigma_e}
\,.\nonumber
\end{eqnarray}

Now we deal with the scalar field $\psi$.  The exterior scalar field
$\psi_e$ is a constant
$\psi_e=\psi_{e_0}$, see Eq.~\eqref{psischw2}, and the
thick shell $\psi$ is also constant $\psi_0$, see Eq.~\eqref{psir}, so
to satisfy the junction condition $\left[\psi\right]=0$, see the
fourth condition in Eq.~(\ref{setstrsmooth}), we put
\be
\psi_{e}=\psi_0\,,\quad\quad r_{\Sigma_e}\leq r<\infty.
\label{psianother}
\ee

The other junction
condition on $\varphi$, namely, $\left[\partial_c\varphi\right]=0$,
see the fifth condition in Eq.~(\ref{setstrsmooth}), follows by
differentiating the solution for $\varphi$ on both sides of
$r_{\Sigma_e}$. From the exterior one has $\partial_r\varphi_{e_0}=0$.
This implies that from the interior one must have
$\partial_r\varphi\left(r_{\Sigma_e}\right)=0$.
The scalar field $\varphi\left(r\right)$ within the thick
shell depends on two constants of integration $\varphi_1$ and
$\varphi_2$. Taking the derivative of $\varphi\left(r\right)$, using
$r=r_{\Sigma_e}=3M$, and forcing the derivative to vanish at
$r_{\Sigma_e}$
so that
the junction condition
$\left[\partial_c\varphi\right]=0$ is obeyed,
one obtains a relationship between $\varphi_1$ and
$\varphi_2$ of the form $\varphi_2=\frac{2}{27}\left(486\pi
M^2\rho_0+\varphi_1\right)$. Inserting these considerations into the
equations for $p_r\left(r\right)$ and $p_t\left(r\right)$, one
verifies that $p_r\left( r_{\Sigma_e} \right)=0$ and $p_t\left(
r_{\Sigma_e} \right)= \frac14\left( -1+\frac{\varphi_1}{54 \pi
M^2\rho_0} \right)\rho_0 $ with $\varphi_1$ a free parameter. Any
$\varphi_1$ is a good choice. In particular,
the choice $\varphi_1=54\pi
M^2 \rho_0$ leads to $p_t\left( r_{\Sigma_e} \right)=0$, which is
the value of the tangential pressure
at $r_{\Sigma_e}$ we assume for
the solution.  For this $\varphi_1$ one gets in addition
$\varphi_2=40\pi
M^2\rho_0$.
Using the values for
$\frac{M}{r_{\Sigma_e}}$, $\varphi_1$, and $\varphi_2$,
just obtained,
means
that the the scalar field $\varphi(r)$ within
the thick shell, see
Eq.~(\ref{varphiofr}), can be written as
$
\varphi\left(r\right)=
\psi_0+
\left[
54
\left(\frac{M}{r}\right)^2+
40\frac{M}{r}-6\left(
\frac{r}{M}\right)^2\right]\pi M^2\rho_0$ for 
$
r_{\Sigma_i}\leq r \leq r_{\Sigma_e}
$, as we have put in 
Eq.~(\ref{shellphi2}).
Thus, this $\varphi\left(r\right)$
guarantees that the junction condition
$\left[\partial_c\varphi\right]=0$, see the fourth condition in
Eq.~(\ref{setstrsmooth}), is satisfied.
Now we can turn
back to the junction condition $\left[\varphi\right]=0$, see
the third condition in Eq.~(\ref{setstrsmooth}).
At the surface, from Eq.~\eqref{shellphi2}, we have
$\varphi\left(r_{\Sigma_e}\right)=\psi_0+
72\pi M^2\rho_0$, where Eq.~\eqref{fracmRe} was used.
So, to satisfy the junction condition $\left[\varphi\right]=0$
one has to impose
$\varphi_{e}=\varphi_{e_0}=\varphi\left(r_{\Sigma_e}\right)$, i.e.,
\be
\varphi_{e}=\psi_0+
72\pi M^2\rho_0\,,\quad\quad r_{\Sigma_e}\leq r<\infty.
\label{varphiext2}
\ee
Note also that from 
Eq.~(\ref{shellphi2}) one finds that at 
$r_{\Sigma_i}$ one obtains 
Eq.~(\ref{varphiinteriorreal})
justifying it.

From
Eq.~(\ref{psianother}), the junction condition
$\left[\partial_c\psi\right]=0$, see the
sixth condition in Eq.~(\ref{setstrsmooth}), is 
trivially satisfied.

For the matter fields one has the energy density
$\rho$ as given in
Eq.~(\ref{rho0}),
$\rho=\rho_0$. 
Also, once $r_{\Sigma_e}=3M$, and $\varphi_1$, and $\varphi_2$
have been found,
the pressures can now be
written as
\begin{eqnarray}
&p_r\left(r\right)=\frac{1}{2}\left(-3-
\frac{27M^4}{r^4}+\frac{10M}{r}\right)\rho_0\,,
\label{profr2}\nonumber\\
&p_t\left(r\right)=\frac{1}{4}\left(-7+
\frac{27M^4}{r^4}+\frac{20M}{r}\right)\rho_0\,,\\
&\quad\quad\quad
r_{\Sigma_i}\leq r \leq r_{\Sigma_e}.\nonumber
\label{ptofr2}
\end{eqnarray}
see
the
Appendix~\ref{calculationsappendixb}.
In the exterior region one has $\rho=0$ and
$p=0$, for $r_{\Sigma_e}\leq r<\infty$.

The values found for the parameters
$\frac{M}{r_{\Sigma_e}}=\frac13$,
$\varphi_1$, $\varphi_2$,
$\varphi_{e_0}$, and
$\zeta_0$,
guarantee that
the matching between the thick shell and the exterior Schwarzschild
solution at $r_{\Sigma_e}$ is smooth.

The final solution has now four free parameters, namely,
$m$ or $r_{\Sigma_i}$, $M$, $\rho_0$ and
$\psi_0$ that can be chosen.  One should put $\psi_0<\varphi$
everywhere, to guarantee that the coupling $\left(\varphi-\psi\right)$
in the field equations Eq.~\eqref{genein2} is positive.

\vskip 1.5cm
\noindent
{\small \it The full solution:}

\vskip 0.1cm
\noindent
The full solution represents a quasistar, i.e., a black hole
surrounded by a spherical nonaccretting
star thick shell, surrounded by
an exterior
vacuum. 
The full solution is then
characterized by the following expressions
and quantities.

For the interior region,
$r_h< r\leq r_{\Sigma_i}$, where
$r_h=2m$,
we have a Schwarzschild
black hole spacetime given by the line element Eq.~(\ref{mink2}) with
the mass parameter ${m{}}$ arbitrary,
the field $\varphi_i$ is a
constant, see Eq.~(\ref{varphimink2}), which
is then found from the junction conditions, see
Eq.~(\ref{varphiinteriorreal}), and 
the field $\psi$ is a constant,
see Eq.~(\ref{psiimink2}), with $\psi_{i}$
then found from the junction conditions, see
Eq.~(\ref{psithinshell}).
Since the interior is vacuum there are no
matter fields, see Eqs.~(\ref{rhomink2}) and~(\ref{pmink2}).

\begin{widetext}

\begin{figure} [h]
\hskip -0.8cm
\includegraphics[scale=0.78]{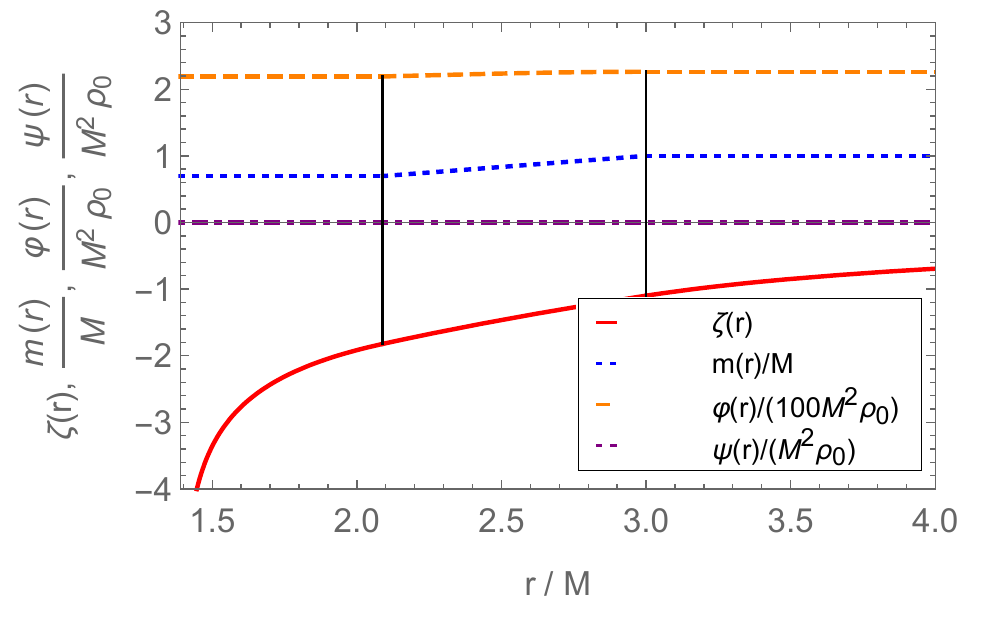}
\includegraphics[scale=0.30]{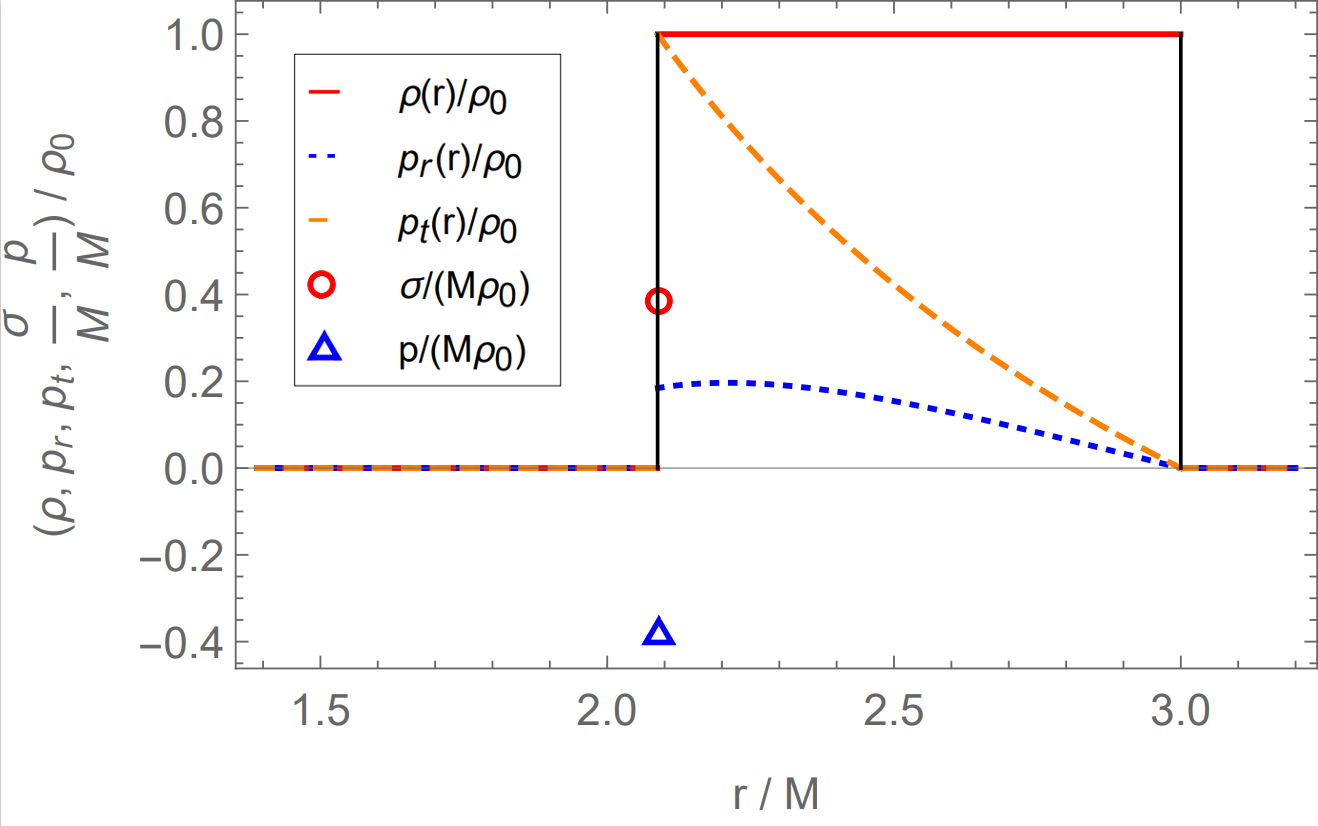}
\caption{
The full solution for the quasistar with a black hole.  In the left
panel it is plotted the metric fields $\zeta(r)$ and $m(r)$ in units
of exterior spacetime mass $M$, and the scalar fields
$\varphi(r)$ and
$\psi(r)$ 
in units of $M^2\rho_0$, which is a dimensionless
quantity, as functions of $r$, more precisely of $\frac{r}M$, and in
the right panel it is plotted the matter fields $\rho(r)$, $p_r(r)$,
$p_t(r)$, $\frac{\sigma}M$, and $\frac{p}M$, in units of $\rho_0$, as
functions of $r$, more precisely of $\frac{r}M$, for the full
solution, i.e., for $r_h< r<\infty$, where $r_h=2m$. Note that
$\sigma$ and $p$ reveal the presence of a thin shell, a mild one, at
$r_{\Sigma_i}$, whereas at $r_{\Sigma_e}$ the solution is smooth.  The
value chosen for $r_{\Sigma_i}$ is $r_{\Sigma_i}=2.09M$.  Since $r_h$
is given by $r_h=\frac23r_{\Sigma_i}$ one has $r_h=1.39M$.  The value
for $r_{\Sigma_e}$ is mandatorily $r_{\Sigma_e}=3M$.  The values
chosen for the other free parameters are $\psi_0=0$, $\rho_0=1$, and
$M=1$, i.e., all is normalized to $M$.  The NEC, WEC, and DEC, are
satisfied for the full solution. See text for further details.
}
\label{fullfields}
\end{figure}

\end{widetext}

For the inferior border of the thick shell one has $r=r_{\Sigma_i}$,
where to satisfy all the energy conditions $r_{\Sigma_i}$ is in the
range given by Eq.~(\ref{rangeRi}) which can be put in terms of $m$
using Eq.~(\ref{sigmai}). The line element can be taken from
Eq.~(\ref{mink2}) evaluated at $r_{\Sigma_i}$, $\varphi$ has the value
given in Eq.~(\ref{varphiinteriorreal}), $\psi$ has the value given in
Eq.~(\ref{psithinshell}), and there is a thin shell with $\sigma$ and
$p$ given in Eqs.~(\ref{sigmashell}) and~(\ref{pshell}), respectively.

For the middle region, i.e., the
thick shell solution, $r_{\Sigma_i}\leq r\leq r_{\Sigma_e}$,
the line element is given in Eq.~(\ref{sphermiddle})
together with Eqs.~(\ref{zetarfinal}) and~(\ref{massfinalfinal}).
In addition,
$\varphi$ is the function given in
Eq.~(\ref{shellphi2}),
$\psi$ has the value given in
Eq.~(\ref{psir}),
and the matter
functions $\rho$, and $p_r$ and $p_t$,
are given in 
Eqs.~(\ref{rho0}), and~(\ref{profr2}), respectively.

For the superior border of the thick shell, one has $r=r_{\Sigma_e}$,
and in terms of $M$ is given by Eq.~(\ref{fracmRe}).  The line element
is given in Eq.~(\ref{schw2}) evaluated at $r_{\Sigma_e}$,
$\varphi$ has the value
given in Eq.~(\ref{varphiext2}),
$\psi$ has
the value given in Eq.~(\ref{psianother}), and there is no thin shell, the
matching is smooth.

For the exterior region, $r_{\Sigma_e}\leq r
<\infty $, we have an exterior Schwarzschild
spacetime given
by the line element Eq.~(\ref{schw2}) with $M$ the spacetime mass,
the field $\varphi$ is a constant given in
Eq.~(\ref{varphiext2}),
the field $\psi$
is a constant given in Eq.~(\ref{psianother}),
and since
it is vacuum there are no matter fields, see
Eqs.~(\ref{rhoschwarz2}) and~(\ref{pschwarz2}).

The full solution is then given by all the equations cited
above.
The full solution is shown
in Fig.~\ref{fullfields}, where plots of
$\zeta(r)$, $m(r)$, $\varphi(r)$, $\psi(r)$,
 $\rho(r)$,
$p_r(r)$,
$p_t(r)$, $\sigma(r_{\Sigma_i})$, and $p(r_{\Sigma_i})$,
are given as function of
the radius $r$.
The radius $r=r_{\Sigma_i}$
can be in the range 
$2.09M\leq r_{\Sigma_i} <3M$, where
$3M=r_{\Sigma_e}$.
For the plots we have chosen
the free parameter $r_{\Sigma_i}$ as 
$r_{\Sigma_i}=2.09M$.
The other free parameters left are chosen as
$\psi_0=0$,
$\rho_0=1$, and the rest is in terms of
$M$, i.e., $M=1$.
All other cases for $r_{\Sigma_i}$ different from
$2.09M$ are similar. The only particular case
worth of note is when
$r_{\Sigma_i}=r_{\Sigma_e}=3M$, in which case
the thick and thin shells disappear, there is no thick
and no thin shell, and the solution
is a Schwarzschild vacuum black hole
with $M=m$.

%

\section{Third application: A wormhole.  Thin shell: Matching a matter
interior to a Schwarzschild-AdS exterior}
\label{thirdapplication}

A third application for the use of junction conditions in the
generalized hybrid metric-Palatini matter theory is to a wormhole
solution.  In the generalized hybrid metric-Palatini matter theory, as
in theories of gravitation in which the gravitational sector is
enlarged, there is the possibility that the energy conditions, in
particular the NEC, for the matter sector are obeyed.  In this manner,
the wormhole is not exotic and the flaring out geometry necessary in
every wormhole solution is supported by the higher-order curvature
terms in the geometrical representation, or by the two fundamental
scalar fields in the scalar-tensor representation.  These terms can
then be interpreted as a gravitational fluid, and in building such a
wormhole one gets exoticity in the gravitational sector from a trade
with exoticity in the matter sector.

We display here a wormhole solution using 
the scalar-tensor representation
of the theory, knowing that in
the geometrical representation we obtain
the same expressions and quantities
for the solution.
The wormhole solution
we want to work out is
composed of three regions.  The first region is the inside region
containing matter, where the wormhole throat is situated.
It has two branches that develop out from the throat
as is usual in a wormhole solution.
The
second region, the middle region,
is composed of a thin shell made of matter,
actually, two similar shells,
that join each interior branch to each exterior part.
The third
region, the exterior
region, is a vacuum Schwarzschild-AdS region that extends up
to infinity.  This wormhole solution
has the important feature
that the NEC for
the matter is verified for the entire spacetime.
Its existence reinforces the believe that
additional fundamental gravitational fields, such as the scalar fields
used here, are behind the construction of wormholes that do not need
exotic matter.  Nonetheless, the engineering of these wormholes is
hard to realize, even theoretically, and so these solutions are
probably scant.
This wormhole solution
has been presented before \cite{rosa2018}.
Here we refer briefly to the solution.
We use consistently
the nomenclature that we have been using
for the metric fields,
the scalar fields, and matter fields. 
The scalar tensor theory that
we employ is one in which the
potential for the
scalar fields  $\varphi(r)$ and $\psi(r)$ 
is $V(\varphi,\psi)$ of the form
$V\left(\varphi,\psi\right)=V_0 \left(\varphi-\psi\right)^2$, where
$V_0$ is some free constant potential.

Let us start to present the
interior region, $r_0\leq r\leq r_\Sigma$. This interior starts
at the wormhole throat with radius $r_0$ and is composed of
two symmetric branches, each branch
is in the range $r_0\leq r \leq r_\Sigma$
and continues in a symmetric manner
up to $r_\Sigma$ where the two branches are then connected
to two
independent exterior regions. 
The metric function $\zeta(r)$ is
given by $\zeta(r)=\zeta_0$, where $\zeta_0$ is a free constant, and
the other metric function $m(r)$ is given by $m(r)=\frac{r_0^2}{r}$,
where $r_0$ is the radius of the wormhole throat, so that the line
element is
$ds^2=-e^{\zeta_0}dt^2+\left(1-\frac{r_0^2}{r^2}\right)^{-1}dr^2
+r^2 d\Omega^2$.
The scalar field $\varphi(r)$ is
given by $\varphi(r)=\psi_0-\frac{r_0^2}{V_0 r^4}$, 
and
the scalar field $\psi(r)$ is $\psi(r)=\psi_0$,
where $\psi_0$ is a constant.  The matter
fields in this interior region, i.e., the energy density $\rho$, the
radial pressure $p_r$, and the tangential pressure $p_t$ of the fluid,
are given by
$\rho= \frac{M^6}{4r^6} \left(\, {\mbox{\scriptsize 31}}
\frac{r_0^2}{r^2} -{\mbox{\scriptsize 24}} \right)\rho_0$, $
p_r=
\frac{M^6}{4r^6}\left(\, {\mbox{\scriptsize 13}}
\frac{r_0^2}{r^2}-{\mbox{\scriptsize 16}} \right)\rho_0$, and
$p_t=
\frac{M^6}{4r^6} \left(
-{\mbox{\scriptsize 39}}\frac{r_0^2}{r^2}
+{\mbox{\scriptsize
32}} \right)\rho_0$, where we have defined for
convenience a standard density $\rho_0$ by $\rho_0\equiv
\frac{r_0^2}{4\pi (-V_0) M^6}$.
Let us now present the solution at the thin shell, i.e., at $r=
r_\Sigma$. There are two thin shells actually, each one
at an $r_\Sigma$ that joins each exterior branch 
of the wormhole solution.
At $r_\Sigma$ the metric function $\zeta(r)$ has the value
$\zeta(r_\Sigma)=\zeta_0$, the metric function $m(r)$ suffers a jump
due to the fact that the shell has some energy-density, actually
negative so that the mass decreases,
the field
$\varphi(r)$ is continuous and thus it has the value
$\varphi(r_\Sigma)= \psi_0-\frac{r_0^2}{V_0 r_\Sigma^4}$, and 
the field $\psi(r)$ is continuous
and thus it has the value $\psi(r_\Sigma)=\psi_0$.
There is
matter in the thin shell, it has surface energy density given by
$\frac{\sigma}{M}= \frac{M^5}{2r_\Sigma^5}\left( \, r_\Sigma
k(r_\Sigma)-4\right)\rho_0$, and surface pressure given by
$\frac{p}{M}= \frac{M^5}{2r_\Sigma^5}\left(4+ \frac12\, r_\Sigma
k(r_\Sigma)\right)\rho_0$, where $k(r_\Sigma)
=\frac{r_0\zeta_0}{2r_\Sigma^2}\sqrt{1- \frac{r_0^2}{r_\Sigma^2}}+
\frac{\frac{r_0^2}{r_\Sigma}-6M}{6r_\Sigma^2
\sqrt{1-\frac{2M}{r_\Sigma}+\frac{r_0^2}{6r_\Sigma^2}}}$, and again
$\rho_0= \frac{r_0^2}{4\pi (-V_0) M^6}$.  The NEC at the thin shell is
obeyed if $\sigma+p>0$.  One has to be careful in choosing $r_\Sigma$
since it has to be greater than the gravitational radius of the
solution, otherwise there would be a horizon and the solution would be
invalid.  For this one has to bear in mind
that the gravitational radius $r_g$ of an
exterior Shwarzschild-AdS spacetime always obeys $r_g<2M$.
Let us continue with the exterior region, $r_\Sigma\leq
r <\infty$.
The exterior region is composed of two symmetric parts
each joining the interior region at
a$r_\Sigma$. 
The
metric function $\zeta(r)$ in the exterior is given by $e^{\zeta(r)}=
\frac{ 1-\frac{2M}{r}-\Lambda r^2 } { 1-\frac{2M}{r_\Sigma}- \Lambda
{r_\Sigma}^2 }e^{\zeta_0} $, where $\Lambda$ is a cosmological
constant given by $\Lambda=\frac{V_0(\varphi_e-\psi_e)}{6}$, and the
metric function $m(r)$ is $m(r)=2M+ \Lambda r^3$, where $M$ is the
exterior spacetime mass, so that the exterior line element is $ds^2=-
\frac{1-\frac{2M}{r}-\Lambda r^2} {1-\frac{2M}{r_\Sigma}-\Lambda
r_\Sigma^2} e^{\zeta_0}dt^2 +\frac{dr^2}{1-\frac{2M}{r}- \Lambda r^2}
+r^2d\Omega^2$, $r_\Sigma\leq r \leq\infty$.
The scalar field $\varphi_e(r)$ is
$\varphi_e(r)=\psi_0-\frac{r_0^2}{V_0 {r_\Sigma}^4}$,
and so is a
constant, and the scalar
field $\psi_e(r)$ is
given by $\psi_e(r)=\psi_0$, also a
constant. Thus, $\Lambda=-\frac{r_0^2}{6r_\Sigma^4}$.
Then, $\Lambda$
is negative and the exterior spacetime is Schwarzschild-AdS.
The matter fields in the exterior
region are given by $\rho=0$, $p_r=0$, and $p_t=0$.

\begin{widetext}

\begin{figure} [h]
\hskip -0.8cm
\includegraphics[scale=0.665]{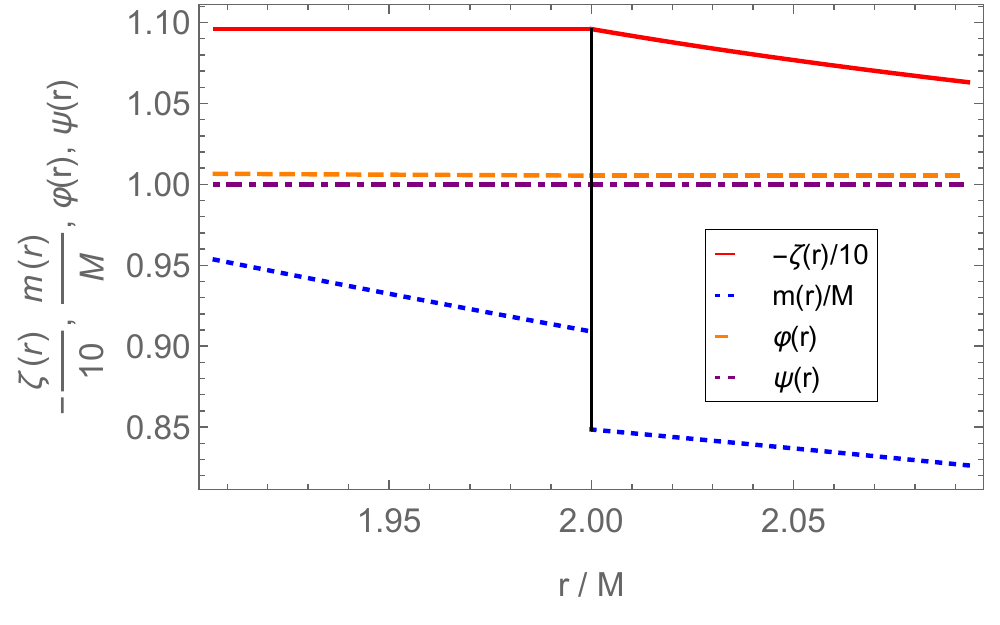}
\includegraphics[scale=0.276]{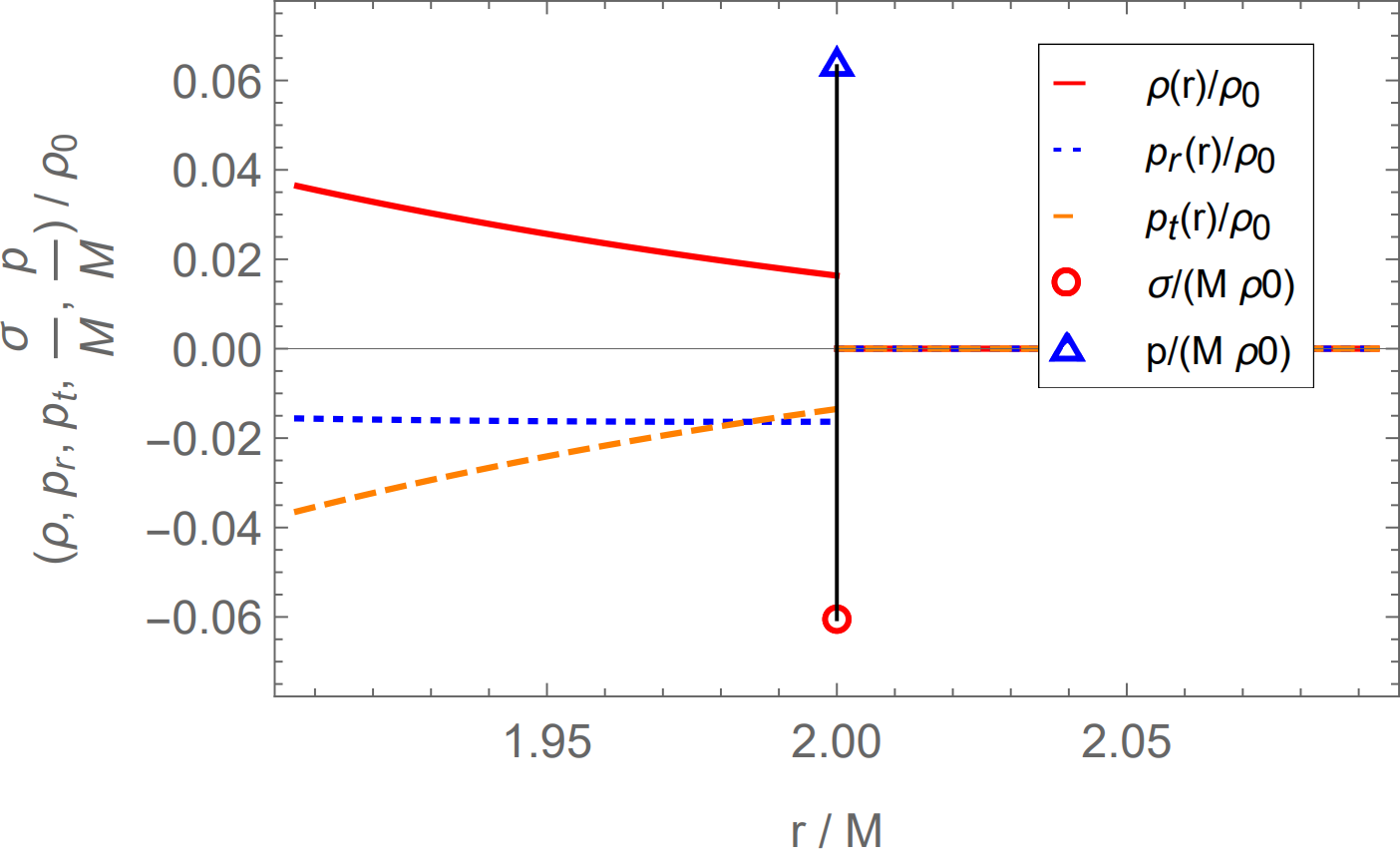}
\caption{
The full solution for the wormhole.  In the left panel it is plotted
the metric fields $\zeta(r)$ and $m(r)$ in units of exterior spacetime
mass $M$, and the scalar fields $\varphi(r)$ and $\psi(r)$ which have
no units, as functions of $r$, more precisely of $\frac{r}M$, and in
the right panel it is plotted the matter fields $\rho(r)$, $p_r(r)$,
$p_t(r)$, $\frac{\sigma}M$, and $\frac{p}M$, in units of $\rho_0$, as
functions of $r$, more precisely of $\frac{r}M$, for the full
solution, i.e., for $r_0< r<\infty$, where $r_0$ is the throat radius
and there are two copies of $r$.  Note that $\sigma$ and $p$ reveal
the presence of thin shell at $r_{\Sigma}$.  The value chosen for the
throat radius $r_0$ is $r_0={\mbox{\scriptsize 2}}
\sqrt{\frac{10}{11}} M$.  The value chosen for $r_{\Sigma}$ is
$r_\Sigma=2M$. The values chosen for the other free parameters are
$\zeta_0=-10.96$, $\psi_0=1$, $V_0=-42$, and $M=1$, i.e., all is
normalized to $M$. Since we have defined $\rho_0= \frac{r_0^2}{4\pi
(-V_0) M^6}$, one gets $\rho_0=0.00689$.
$\Lambda=-\frac{r_0^2}{6r_\Sigma^4}$, so $\Lambda=-0.03788$.  Also,
$\frac{\sigma}M=-0.0605\rho_0$, and $\frac{p}M=0.0632\rho_0$.
The matter NEC is
satisfied for the full wormhole solution.
See text for further details. 
}
\label{wormsol2}
\end{figure}

\end{widetext}

To have a definite solution we choose
$r_0={\mbox{\scriptsize 2}}
\sqrt{\frac{10}{11}}
M$, $r_\Sigma=2M$, $\zeta_0=-10.96$,
$\psi_0=1$, and $V_0=-42$.  The choice for a negative $V_0$ is that it
makes the quantity $\sigma+p$ positive and so the NEC is satisfied.
Indeed, $\sigma=-0.0605\rho_0M$ and $p=0.0632\rho_0M$, so that
$\sigma+p=0.0027\rho_0M$, is positive, and the matter in the shell
obeys the NEC.
$\Lambda$ is negative, it is
$\Lambda=-0.03788$, thus the exterior is a
Schwarzschild-AdS spacetime.  We normalize all quantities to the
spacetime mass $M$, which amounts to put $M=1$.  The matter WEC does
not hold on the shell but holds everywhere else.  The matter NEC,
which is the most important energy condition, is obeyed everywhere in
conformity with our aim.  This completes our full wormhole solution.
Figure~\ref{wormsol2} displays the solution.  For a full account of
this wormhole solution see \cite{rosa2018}.

\vskip 0.1cm

\vskip 0.1cm





\section{Conclusions}\label{conclusion}

The generalization from the
general relativity action containing the Ricci scalar $R$ alone plus
matter to generalized hybrid metric-Palatini gravity action containing
in the geometric representation a function of the Ricci scalars $R$
and $\cal R$, and in the scalar-tensor representation a function of
the Ricci scalar $R$ and two scalar fields $\psi$ and $\phi$, in which
one can add matter fields to both, leads to new junction conditions in
both representations.
We have presented the junction conditions for the generalized hybrid
metric-Palatini gravity, for both the geometrical and the
scalar-tensor representation of the theory, in the cases of a thin
shell junction and of a smooth matching.

In the geometrical representation of the generalized hybrid
metric-Palatini gravity, in the case of a
matching with a thin shell, the conformal relation between $R$ and
$\mathcal R$, implies that the jump of the derivatives of $R$ and
$\mathcal R$ do not contribute independently to the stress energy
tensor of the shell. In fact, the two quantities must have a specific
relation.  It also follows that the normal derivative
of the Ricci scalar being not continuous, gives rise to the existence
of the matter thin shell.  In the case of a smooth matching, we
obtained that the trace of the extrinsic curvature and the Ricci
scalar must be continuous, and also that the normal derivative of the
Ricci scalar must be continuous.  Moreover, given the conformal
relation between $R$ and $\mathcal R$, it follows that the continuity
of $R$ implies the continuity of $\mathcal R$.

In the scalar-tensor representation of the generalized hybrid
metric-Palatini gravity, some interesting results also appeared which
can be compared with the junction conditions for the Brans-Dicke
theory.  In the case of a matching with a thin shell, the trace of the
extrinsic curvature must be continuous, on one hand emphasizing the
relation between the geometrical and the scalar-tensor representation
of the theory, on the other hand showing its difference to the
Brans-Dicke theory where the trace of the extrinsic curvature does not
need to be continuous.  Moreover, since only the scalar field
$\varphi$ is coupled to matter, only the derivative of $\varphi$ is
allowed to be discontinuous in which case it contributes to the stress
energy of the thin shell.  In the case of a smooth matching, it turns
out that the scalar fields and their normal derivatives must be
continuous.  As well, the trace of the extrinsic curvature must be
continuous.

The importance and usefulness of these junction conditions is
explicit, as it helped to find three different solutions, one for a
spacetime with a thin shell as a simplified model of a star, other for
a three-region spacetime with a thick shell as a simplified model of a
quasistar, and yet another for a wormhole.

The use of the junction conditions allowed us to
satisfy all the energy conditions for the star thin shell and the
quasistar thick shell solutions, and to satisfy the matter NEC for the
wormhole spacetime solution which generically is something very hard
to achieve.  These results point towards the physical relevance of the
solutions obtained.  Of course, to progress further with the physical
meaningfulness of the solutions found, one has to perform a stability
study to radial perturbations to start with, and then a full stability
analysis.

It should be mentioned that,
although the equivalence between the two representations, geometrical
and scalar-tensor, can be unambiguous in some results, such as the
continuity of the trace of the extrinsic curvature or the direct
implication between the continuity of the scalar fields to the
continuity of $R$ and $\mathcal R$, the equivalence of the remaining
results is not so clear cut.  For instance, in the second application
of a quasistar which has a thick shell, although not obvious and not
made explicitly, one can show that the transformation $f_R=\varphi$
yields the same expression in both representations, geometric and
scalar-tensor, in what concerns the density and transverse pressure of
the inner thin shell.  Moreover, the relation between the derivatives
of $R$ and $\mathcal R$ in the geometrical representation can also be
found in the scalar tensor representation since a replacement
$f_\mathcal R=\psi$ leads to the correct relation.  We may then
conclude that the junction conditions in both representations are
indeed equivalent.

It has been also
made clear that matched solutions in the generalized
hybrid metric-Palatini gravity have more restrictions than matched
solutions in general relativity.
This means that matched solutions in
general relativity may not necessarily be solutions in the generalized
hybrid metric-Palatini gravity, a simple example of this fact is for
self-gravitating thin shells of matter that in the generalized hybrid
metric-Palatini gravity are constrained to have a specific radius
$r_\Sigma=\frac{9M}4$, whereas in general relativity  the shells can
have any radius $r_\Sigma$ as long as $r_\Sigma$
is greater than the gravitational radius.

The restriction on the matching radius 
arises from the extra junction condition $\left[K\right]=0$, which
does not exist
in general relativity.  This extra junction condition is common
in metric theories of gravity where the Lagrangian has an arbitrary
dependence on the metric Ricci scalar $R$, from which
$f\left(R\right)$ and $f\left(R,T\right)$ are other known examples. To
avoid the $\left[K\right]=0$ junction condition,
one must recur to metric-affine theories of gravity, a
well-known example being the case of the pure Palatini
$f\left(\mathcal R\right)$ theory, and another possible example being
the case of the Palatini $f\left(\mathcal R,T\right)$ gravity theory.

\begin{acknowledgments}
We thank Gonzalo Olmo for discussions and suggestions. JLR is
supported by the European Regional Development Fund and the program
Mobilitas Pluss, Project No.~MOBJD647. JPSL thanks Funda\c{c}\~{a}o
para a Ci\^{e}ncia e Tecnologia FCT - Portugal for financial support
through Project No.~UIDB/00099/2020.
\end{acknowledgments}



\appendix

\section{Explicit equation of
Sec.~\ref{equations1}}
\label{longformula}

We have derived the field equations of the generalized hybrid
metric-Palatini gravity in the geometrical representation in terms of
derivatives of the function $f\left(R,\mathcal R\right)$ in
Sec.~\ref{equations1}. As shown in
Eq.~\eqref{partialf}, these derivatives can be expanded in terms of
derivatives of $R$ and $\mathcal R$. Since the
complete expanded field equation 
is big, we opted to not write it there,
we do it here. For that, we insert the set of
equations given in Eq.~\eqref{partialf} into the field equation given
in Eq.~\eqref{genfield2}, and obtain the fully extended field
equation in the form
\begin{widetext}
\begin{eqnarray}
&&R_{ab}\left(f_R+f_\mathcal R\right)-\left(f_{RR}+f_{\mathcal R
R}\right)\nabla_a\nabla_b-\left(f_{R\mathcal R}+f_{\mathcal R\mathcal
R}\right)\nabla_a\nabla_b\mathcal R-\left(f_{RRR}+f_{R\mathcal
R\mathcal R}\right)\partial_a R\partial_b R-\left(f_{R\mathcal
R\mathcal R}+f_{\mathcal R\mathcal R\mathcal
R}\right)\partial_a\mathcal R\partial_b\mathcal R-\nonumber \\
&&-2\left(f_{R\mathcal R R}+f_{\mathcal R\mathcal R
R}\right)\partial_{(a}R\partial_{b)}\mathcal
R+g_{ab}\left[\left(f_{RR}-\frac{1}{2}f_{\mathcal R R}\right)\Box
R+\left(f_{R\mathcal R}-\frac{1}{2}f_{\mathcal R\mathcal
R}\right)\Box\mathcal R+\left(f_{RRR}-\frac{1}{2}f_{R\mathcal
R\mathcal R}\right)\partial_cR\partial^c R+\right.\nonumber \\
&&+\left.\left(f_{R\mathcal R\mathcal R}-\frac{1}{2}f_{\mathcal
R\mathcal R\mathcal R}\right)\partial_c\mathcal R\partial^c\mathcal
R+\left(2f_{R\mathcal R R}-f_{\mathcal R\mathcal R R}\right)\partial_c
R\partial^c\mathcal R\right]+\frac{3}{2f_\mathcal R}\left(f_{\mathcal
R R}^2\partial_aR\partial_b R+f_{\mathcal R\mathcal
R}^2\partial_a\mathcal R\partial_b\mathcal R+2f_{\mathcal R
R}f_{\mathcal R\mathcal R}\partial_{(a}R\partial_{b)}\mathcal
R\right)-\nonumber \\ &&-\frac{1}{3}g_{ab}f=8\pi T_{ab}.
\end{eqnarray}
\end{widetext}

\section{A more general theory for the star shell application of
Sec.~\ref{app} }
\label{choosingatheory}

In Sec.~\ref{app}, in the first application of the junction conditions
formalism in generalized hybrid metric-Palatini gravity, on a star
thin shell matching an interior Minkowski spacetime to an exterior
Schwarzschild spacetime, we have chosen a specific theory for
$f\left(R,\mathcal R\right)$, namely, $f\left(R,\mathcal R\right) =
R+\mathcal R+\frac{R \mathcal R}{R_0}$.  Here we give a more general
$f\left(R,\mathcal R\right)$ and find the matter properties for the
thin shell in the geometrical representation.

We consider the following form of the
function $f\left(R,\mathcal R\right)$,
\be\label{specf}
f\left(R,\mathcal R\right)=g\left(R\right)+
\mathcal R h\left(\dfrac{R}{R_0}\right)
\ee
where $g\left(R\right)$ and
$ h\left(\dfrac{R}{R_0}\right)$ are well behaved
functions of their arguments, and $R_0$ is a constant with units of
$R$. For this specific choice of the function $f$, we can write the
derivatives $f_R$ and $f_\mathcal R$ as
\be\label{specderiv}
f_R=g'\left(R\right)+h'\left(\bar R\right)\bar{\mathcal R},
\quad\quad f_\mathcal R=h\left(\bar R\right),
\ee
and the second derivatives $f_{RR}$, $f_{R\mathcal R}$
and $f_{\mathcal R\mathcal R}$ as
\be
\label{secondderiv}
f_{RR}=g''\left(R\right)+h''\left(\bar R\right)
\frac{\bar{\mathcal R}}{R_0},
\; f_{{\mathcal R}R}=h'\left(\bar R\right)\frac{1}{R_0},
\;
f_{{\mathcal R}{\mathcal R}}=0\,,
\ee
where $\bar{\mathcal R}= \frac{\mathcal R}{R_0}$
and $\bar{R}=\frac{R}{R_0}$
are 
dimensionless variables. Eq.~\eqref{riccirel} becomes an equation
for $\mathcal R_{ab}$ as a function of $R_{ab}$ and $R$ as
\beq\label{specrel}
\mathcal R_{ab}=&&R_{ab}-\frac{1}{h\left(\bar R\right)}
\left(\nabla_a\nabla_b+\frac{1}{2}g_{ab}\Box\right)h
\left(\bar R\right)+\nonumber\\
&&+\frac{3}{2 h\left(\bar R\right)^2}\partial_ah
\left(\bar R\right)\partial_bh\left(\bar R\right).
\eeq
Notice that the specific choice of the function $f$
in Eq.~\eqref{specf} allows us to write $\mathcal R_{ab}$
as a function of $R_{ab}$ and $R$ only, implying that we can use
Eq.~\eqref{specrel} and its trace, and use Eqs.~\eqref{specf}
and~\eqref{specderiv} to cancel the terms depending on $\mathcal
R_{ab}$ and $\mathcal R$ in the
field equation given in
Eq.~\eqref{genfield2} and obtain an equation that only depends on the
metric $g_{ab}$ and its derivatives.
The simplification presented
in Eq.~\eqref{specrel}
is not possible in general.
Indeed, for a generic choice of the
function $f$ for which $f_\mathcal R$ depends on $\mathcal R$,
Eq.~\eqref{riccirel} becomes a partial differential equation for
$\mathcal R$ and the problem is much more complicated.

Considering a Minkowski spacetime inside and a Schwarzschild
spacetime outside, we can find the
matter properties of the shell at $r_\Sigma$ for
the theory
given in Eq.~\eqref{specf}. 
All we have worked out in Sec.~\ref{app} for the geometrical
representation follows apart the the sixth
junction condition of  Eq.~(\ref{fullsetshell}).
The 
two components of  Eq.~(\ref{fullsetshell})
correspond to $S_0^0=-\sigma$ and $S_\theta^\theta=p$.
Since,
the the second junction condition in Eq.~\eqref{fullsetshell} is
$\left[K\right]=0$, for a spherically symmetric thin shell one
can write $\left[K_0^0\right]=-2\left[K_\theta^\theta\right]$,
and so
the sixth
junction condition of  Eq.~(\ref{fullsetshell})
is given by the two independent components,
namely,
$\frac{\epsilon}{8\pi}\left[\left(f_R+f_\mathcal R\right)
\left[K_0^0\right]-n^c\left[\partial_cR\right]\left(f_{RR}-
\frac{f_{\mathcal R R}^2}{f_{\mathcal R\mathcal R}}\right)\right]
=\sigma$
and
$\frac{\epsilon}{8\pi}\left[n^c\left[\partial_cR\right]
\left(f_{RR}-\frac{f_{\mathcal R
R}^2}{f_{\mathcal R\mathcal R}}\right)+
\frac{1}{2}\left(f_R+f_\mathcal R\right)\left[K_0^0\right]\right]=p$,
see Eqs.~(\ref{geos}) and~(\ref{geop}). 
From these two equations
we
obtain for our specific case, i.e.,
from 
Eqs.~\eqref{specf},~\eqref{specderiv}, and~\eqref{secondderiv},
and using $\epsilon=1$ since $n^a$ points in the radial
direction and thus is a spacelike vector, 
\be
\sigma=\frac{1}{8\pi}\left\{\left[g'
\left(R\right)+h\left(\bar R\right)\right]
\frac{16}{27M}\right\}\,,
\label{s1anew}
\ee
\be
p=\frac{1}{16\pi}\left\{\left[g'
\left(R\right)+h\left(\bar R\right)\right]
\frac{16}{27M}\right\}\,,
\label{p1anew}
\ee
where we have used that
since 
$\left[K_0^0\right]=\frac{M}{r_\Sigma^2}
\frac1{\sqrt{1-\frac{2M}{r_\Sigma}}}$
and $r_\Sigma = \frac{9}{4}M$,
see Eq.~\eqref{cond},
one has 
$\left[K_0^0\right]|_{r_\Sigma}=\frac{16}{27M}$.
Thus, from these equations,
one has $\sigma=2p$, which is the expected result 
as the matching is being performed at the Buchdahl
radius, as stated before.

To have a direct comparison
of the geometrical representation results
with the scalar-tensor representation
results
one has to specify form for 
$f\left(R,\mathcal
R\right)$. It is at this point that we choose
$g\left(R\right)$ and $h\left(\bar
R\right)$ as
\be\label{particular1}
g\left(R\right)=R, \qquad h
\left(\frac{R}{R_0}\right)=1+\frac{R}{R_0},
\ee
for which the function
$f\left(R,\mathcal R\right)$ is
\be\label{partfanew}
f\left(R,\mathcal R\right) =
R+\mathcal R+\frac{R \mathcal R}{R_0}\,,
\ee
the one given in Eq.~\eqref{partf}.
For this particular form of the function $f$,
one has 
$g'\left(R\right)=1$ and $h\left(\bar R\right)=1$,
and 
the first and second
derivatives $f_R$, $f_\mathcal{R}$, $f_{RR}$, $f_{\mathcal R\mathcal
R}$ and $f_{R\mathcal R}$ become
$
f_R=1+\bar{\mathcal R}$,  $f_\mathcal R=1+\bar{R}$,
$f_{RR}=f_{\mathcal R\mathcal R}=0$,
$f_{R\mathcal R}=\frac{1}{R_0}$.
Then, 
Eqs.~\eqref{s1anew} and ~\eqref{p1anew}
give
$\sigma=\frac{4}{27\pi M}$ and $
p=\frac{2}{27\pi M}$, respectively, 
which are Eqs.~(\ref{specS}) and (\ref{specSp}) of Sec.~\ref{app}.


\begin{widetext}
\section{Solution for the radial and tangential pressures of the thick
shell of Sec.~\ref{secondapplication}}
\label{calculationsappendixb}
\renewcommand\thefigure{\thesection\arabic{figure}}
\setcounter{figure}{0}

In Sec.~\ref{secondapplication} we have seen
that the matter energy density $\rho$ of the thick shell is a constant 
$\rho=\rho_0$. Before performing the
matching, the radial and tangential pressures
on the thick shell
presented succinctly in Eq.~(\ref{profr})
are here displayed explicitly.
They are
\begin{eqnarray}
&p_r\left(r\right)=
\frac{1}{64\pi r_{\Sigma_e}\left(7M-3r_{\Sigma_e}\right)
r^{\frac{1}{2}\left(5+\beta
\right)}}
\left\{\left(r_{\Sigma_e}\right)^2\left[32\pi r^{
\frac{5+\beta}{2}}\left(1+\beta\right)\rho_0+3\left(7+
\beta^2\right)\varphi_1-3r^\beta\left(\beta^2+6\beta-7
\right)\varphi_2\right]\right.\nonumber
\\
&+14M^2\Big[\left(3+\beta^2\right)\varphi_1-
r^\beta\left(\beta^2+6\beta-3\right)\varphi_2\Big]-
Mr_{\Sigma_e}\left[32\pi r^{\frac{5+\beta}{2}}\left(3+
2\beta\right)\rho_0+\left(67+13\beta^2\right)\varphi_1
\right.\nonumber
\\
&\left.\left.
-
r^\beta\left(13\beta^2+78\beta-67\right)\varphi_2
\right.\right.
\Big]\Big\}\,,
\nonumber\\
&p_t\left(r\right)=-
\frac{1}{64\pi r_{\Sigma_e}\left(7M-3r_{\Sigma_e}\right)
r^{\frac{1}{2}\left(5+\beta
\right)}}
\left[8\pi r^{r_{\Sigma_e}\frac{5+\beta}{2}}\left(\beta^2+
2\beta+13\right)\rho_0+2\left(7M-3r_{\Sigma_e}\right)
\left(3+\beta^2\right)\varphi_1\right.\nonumber
\\
&
\left.
+6r^\beta\left(7M-
3r_{\Sigma_e}\right)\left(\beta-1\right)^2\varphi_2\right]
\,,\quad\quad\quad r_{\Sigma_i}\leq r \leq r_{\Sigma_e}\,,
\end{eqnarray}
where
$\beta=\sqrt{\frac{r_{\Sigma_e}+6 M}{r_{\Sigma_e}-2M}}$,
$\rho_0$ is the constant energy density of the
thick shell,
$M$ is the exterior spacetime mass,
$r_{\Sigma_e}$ is the exterior radius of the thick shell,
and $\varphi_1$
and $\varphi_2$ are constants of integration
that appear is the solution for $\varphi(r)$, see
Eq.~(\ref{varphiinteriorreal}).
For the values obtained using the junction conditions, i.e.,
$r_{\Sigma_e}=3M$,
$\varphi_1=54\pi
M^2 \rho_0$, and 
$\varphi_2=40\pi
M^2\rho_0$, one finds
$p_r\left(r\right)=\frac{1}{2}\left(-3-
\frac{27M^4}{r^4}+\frac{10M}{r}\right)\rho_0$
and
$p_t\left(r\right)=\frac{1}{4}\left(-7+
\frac{27M^4}{r^4}+\frac{20M}{r}\right)\rho_0$,
which are the expressions given in Eq.~(\ref{ptofr2}).

The unprocessed solution for the middle region, i.e., the thick shell
solution without taking into account the junction conditions, is
visualized in Fig.~\ref{fieldstshell}, where all the fields are
plotted.

\begin{figure}[h]
\hskip -1.0cm
\includegraphics[scale=0.58]{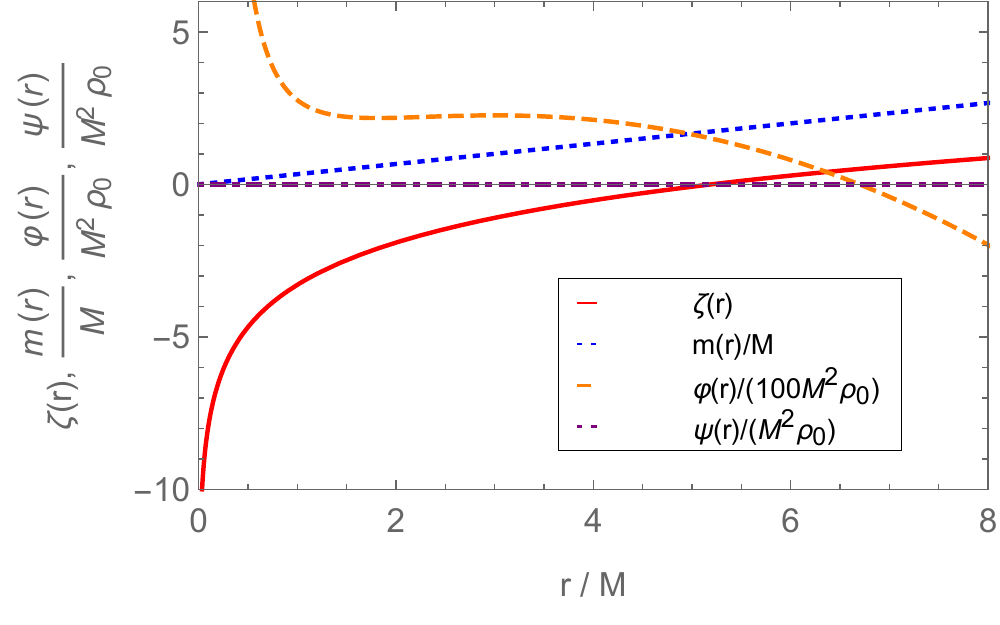}
\includegraphics[scale=0.60]{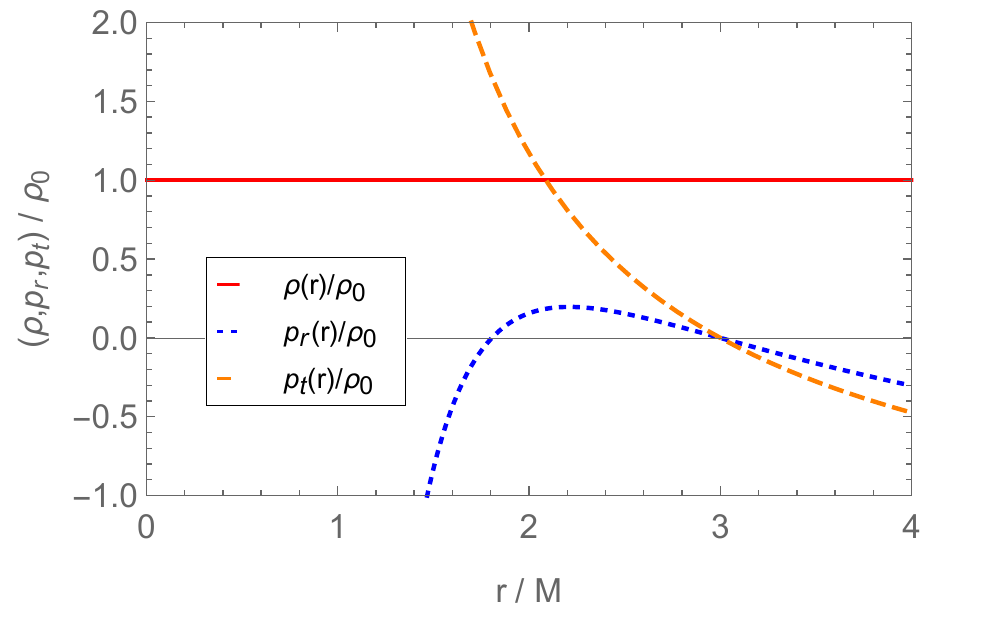}
\caption{
The unprocessed solution for the thick shell.  In the left panel it is
plotted the metric fields $\zeta(r)$ and $m(r)$ in units of exterior
spacetime mass $M$, and the scalar fields $\varphi(r)$
and $\psi(r)$ 
in units of $M^2\rho_0$, which is a dimensionless quantity, as
functions of $r$, more precisely of $\frac{r}M$, and in the right
panel it is plotted the matter fields $\rho(r)$, $p_r(r)$, $p_t(r)$,
in units of $\rho_0$, as functions of $r$, more precisely of
$\frac{r}M$, valid for $0\leq r<\infty$.  The values chosen for the
free parameters are $\psi_0=0$, $\rho_0=1$, and $M=1$, i.e., all is
normalized to $M$.
}
\label{fieldstshell}
\end{figure}

\end{widetext}

\centerline{}
\newpage

\centerline{}
\newpage

\end{document}